%% file: main.tex
\documentclass[onecolumn,11pt]{article}

\usepackage[letterpaper,margin=1in]{geometry}
\usepackage{times}

\input{includes}
\input{macros}

\title{First-Order Logic for Flow-Limited Authorization}

\newcommand{\email}[1]{\href{mailto:#1}{\upshape\detokenize{#1}}}
\author{%
  Andrew K. Hirsch \\[-1pt]
  {\normalsize MPI-SWS\thanks{Work done while author was at Cornell University}} \\[-1pt]
  {\normalsize \email{akhirsch@mpi-sws.org}}
  \and
  Pedro H. Azevedo de Amorim \\[-1pt]
  {\normalsize Cornell University} \\[-1pt]
  {\normalsize \email{pamorim@cs.cornell.edu}}
  \and
  Ethan Cecchetti \\[-1pt]
  {\normalsize Cornell University} \\[-1pt]
  {\normalsize \email{ethan@cs.cornell.edu}}
  \and
  Ross Tate \\[-1pt]
  {\normalsize Cornell University} \\[-1pt]
  {\normalsize \email{ross@cs.cornell.edu}}
  \and
  Owen Arden \\[-1pt]
  {\normalsize University of California, Santa Cruz} \\[-1pt]
  {\normalsize \email{owen@soe.ucsc.edu}}}
\date{}

\begin{document}

\maketitle

\begin{abstract}
  \input{abstract}
\end{abstract}

\input{intro}
\input{by_example}
\input{using_flafol}
\input{proof_system}
\input{proof_theory}
\input{non-interference}
\input{future}
\input{related}
\input{conclusion}

\input{acknowledgements}

\bibliographystyle{alpha}
\bibliography{bibtex/pm-master}

\appendix
\input{examples_flafol}
\input{permissionmodels}

\input{coq}
\input{fullproofsystem}
\input{CompatibleSupercontexts}
\end{document}

%% file: includes.tex
\usepackage{fixltx2e}
\usepackage{graphicx}
\usepackage{grffile}
\usepackage{longtable}
\usepackage{wrapfig}
\usepackage{rotating}
\usepackage[normalem]{ulem}
\usepackage{amsmath}
\usepackage{mathtools}
\usepackage{textcomp}
\usepackage{amssymb}
\usepackage{capt-of}
\usepackage{mathpartir}
\usepackage{stmaryrd}
\usepackage{amsthm}
\usepackage{enumitem}
\setlist{topsep=0.5em,parsep=0pt,partopsep=0pt,itemsep=0.33em}
\usepackage{xparse}
\usepackage{mathdots}
\usepackage{suffix}
\usepackage{xspace}
\usepackage{microtype}
\usepackage{tabularx}
\usepackage{cancel}
\usepackage{multicol}
\usepackage[hidelinks]{hyperref}
\usepackage{cite}



%% file: macros.tex
\renewcommand{\paragraph}[1]{\vspace{0.5\baselineskip}\noindent\textbf{\upshape \mbox{#1.}}\xspace\hspace{0.1em}}

\newcommand{\pageOfRulesSize}{\small}

\newcommand{\synfont}[1]{\text{\sffamily\upshape{#1}}}
\newcommand{\saysfont}[1]{\text{\sffamily\upshape{#1}}}

\newcommand{\proves}{\mathrel{\vdash}}
\newcommand{\nproves}{\mathrel{\nvdash}}

\newcommand{\True}{\synfont{True}}
\newcommand{\False}{\synfont{False}}
\newcommand{\says}{\mathrel\saysfont{says}}
\WithSuffix\newcommand\says*[1]{\says_{#1}}
\newcommand{\canr}{\synfont{CanRead}}
\newcommand{\canw}{\synfont{CanWrite}}
\newcommand{\spksfr}{\mathrel\synfont{SF}}

\newcommand{\influences}{\mathrel{\synfont{CanInfl}}}

\newcommand{\flowsto}{\sqsubseteq}

\newcommand{\labl}{\synfont{Label}}
\newcommand{\prin}{\synfont{Principal}}
\newcommand{\conf}{\synfont{Confidentiality}}
\newcommand{\integ}{\synfont{Integrity}}
\newcommand{\tkn}{\synfont{Token}}

\newcommand{\atl}{\mathrel{@}}

\newcommand{\belief}[2]{#1 \atl #2}
\newcommand{\impl}{\rightarrow}
\newcommand{\mimpl}[1]{\impl}
\WithSuffix\newcommand\mimpl*{\mimpl{\ell}}
\newcommand{\subst}[3]{{#1}[{#2} \mapsto {#3}]}

\newcommand{\binder}[4]{{#1} {#2} \mkern-3mu : \mkern-3mu {#3}.\mkern3mu {#4}}
\newcommand{\forallexp}[3]{\binder{\forall}{#1}{#2}{#3}}
\WithSuffix\newcommand\forallexp*[1]{\forallexp{x}{\sigma}{#1}}
\newcommand{\existsexp}[3]{\binder{\exists}{#1}{#2}{#3}}
\WithSuffix\newcommand\existsexp*[1]{\existsexp{x}{\sigma}{#1}}

\newcommand{\maybemarked}[1]{\overset{\circ}{#1}}
\WithSuffix\newcommand\maybemarked*[2]{\overset{\raisebox{.5pt}{\textcircled{\raisebox{-.9pt}{#1}}}}{#2}}

\newcommand{\fv}{\mathord{\textsf{FV}}}

\newcommand{\modal}[2]{{#1}\langle {#2} \rangle}
\newcommand{\modalG}[1]{\modal{}{}}

\newcommand{\inmodals}{\mathcal{G}}

\newcommand{\isFriend}{\synfont{IsFriend}}

\newcommand{\Alice}{\synfont{Alice}}
\newcommand{\Bob}{\synfont{Bob}}
\newcommand{\Cathy}{\synfont{Cathy}}

\newcommand{\pub}{\synfont{Public}}

\newcommand{\secret}{\synfont{Secret}}
\newcommand{\topsec}{\synfont{TopSecret}}
\newcommand{\frnd}{\synfont{Friends}}

\newcommand{\Input}{\synfont{DBInput}}
\newcommand{\untrusted}{\synfont{LInt}}
\newcommand{\trusted}{\synfont{HInt}}
\newcommand{\sys}{\synfont{System}}

\newcommand{\san}{\synfont{San}}
\newcommand{\tknconf}{\synfont{TknConf}}
\newcommand{\tkninteg}{\synfont{IntegOfTkn}}
\newcommand{\projconf}{\ensuremath{\pi_C}}
\newcommand{\projinteg}{\ensuremath{\pi_I}}
\newcommand{\integof}{\synfont{IntegOf}}

\newcommand{\rootofauth}{\synfont{RootOfAuth}}
\newcommand{\hastkn}{\synfont{HasToken}}

\newcommand{\supdag}{\textsuperscript{\textdagger}}

\newenvironment{syntax}
{%
  \newcommand{\emptysynclass}[2]{\text{##1} & ##2 & & \\}%
  \newcommand{\synclass}[3]{\text{##1} & ##2 & ::= & ##3\\}%
  \newcommand{\altclause}[1] {%
    & & \mid & ##1\\%
  }%
  \newcommand{\alt}{\,\mid\,}%
  \begin{array}{crcl}%
}
{\end{array}}

\newtheorem{thm}{Theorem}

\newtheorem{cor}{Corollary}

\theoremstyle{definition}


%% file: abstract.tex
We present the Flow-Limited Authorization First-Order Logic~(FLAFOL), a logic for reasoning about authorization decisions in the presence of information-flow policies.
We formalize the FLAFOL proof system, characterize its proof-theoretic properties, and develop its security guarantees.
In particular, FLAFOL is the first logic to provide a non-interference guarantee while supporting all connectives of first-order logic.
Furthermore, this guarantee is the first to combine the notions of non-interference from both authorization logic and information-flow systems.
All theorems in this paper are proven in Coq.


%% file: intro.tex
\section{Introduction}
\label{sec:introduction}

Distributed systems often make authorization decisions based on private data, which a public decision might leak.
Preventing such leakage requires nontrivial reasoning about the interaction between information flow and authorization policies~\cite{secpal-plus,flam,flac}.
In particular, the justification for an authorization decision can violate information-flow policies.
To understand this concern, consider a social network where
Bob can say that only his friends may view his photos, and that furthermore only his friends may know the contents of his friend list.
If Alice is not on Bob's friend list and she is denied access to one of his photos, the denial leaks Bob's private information: that Alice is not on Bob's friend list.
Worse, if Alice can indirectly determine what other principals are permitted to see Bob's photos, she could completely enumerate the friend list.

Reasoning about the interaction between information flow and authorization policies is challenging for several reasons.
First, authorization logics and information-flow systems use different notions of trust.
Information-flow systems tend to focus on tracking data dependencies by representing information-security policies as \emph{labels} on data.
They then represent trust as a \emph{flows-to} relation between labels, which determines when one piece of data may safely influence another.
In contrast, authorization logics tend to directly encode \emph{delegations} between principals as a \emph{speaks-for} relation.
Such delegations are often all-or-nothing, where a delegating principal trusts any statements made by the trusted principal, although some logics~(e.g., \cite{hk00spki,SecPAL,NAL}) support restricting delegations to specific statements.
Flows-to relations implicitly encode delegations while speaks-for relations implicitly encode permitted flows.
To understand \emph{how}, we must understand how these disparate notions of trust interact.

Both forms of trust serve to selectively constrain the communication that system components rely on to make secure authorization decisions.
For example, in the social network example above, suppose Bob's security settings are recorded on server~$X$, and his photos are stored on server~$Y$.
When Alice tries to view Bob's photo, server~$Y$ communicates with server~$X$ to determine if Alice is permitted to do so.
Modeling this communication is important because (1) the servers that $Y$ communicates with influence its authorization decisions, and (2) 
communication can leak private information.

Describing the information security of authorization decisions such as the one above requires modifying typical authorization policies to include information flow.
Information-flow systems are excellent at tracking when and what information one principal communicates to another, specifically by transferring data from one label to another.
It is less clear when communications occur in authorization logics.
A common approach~\cite{NAL,labw91,abadi06} simply models Alice delegating trust to Bob as Alice importing all of Bob's beliefs.

Authorization logics \emph{do}, however, excel at reasoning about beliefs.
Authorization logics allow us to write $\Alice \says \varphi$, meaning that Alice believes formula~$\varphi$.
This $\says$ statement is itself a formula, so we can reason about what Bob believes Alice believes by nesting $\says$ formulae.
Information flow, in contrast, has no notion of belief, and so cannot reason about principals' beliefs about each others' beliefs.

In order to express authorization policies, not only does one need the ability to express trust and communication, but also a battery of propositions and logical connectives.
Any tool that combines authorization and information flow should be capable of expressing enough logical connectives to reason about
real-world policies.
First-order logic seems to be a sweet spot of expressive power: it can encode most authorization policies, but it is still simple enough to have clean semantics.
For instance, Nexus~\cite{NAL,sirer11}---a distributed operating system that uses authorization logic directly in its authorization mechanisms---
can encode all of its authorization policies using first-order logic.
\footnote{The Nexus Authorization Logic is actually a monadic second-order logic, but this is used only to encode $\textsf{speaksfor}$; their examples only use first-order quantification~\cite{NAL}.}

Finally, evaluating any attempt to combine authorization and information flow policies must examine the resulting security guarantees.
Both authorization logics and information-flow systems have a security property called \emph{non-interference}.
Information-flow systems view non-interference as standard, while authorization logics often view it as desirable but unobtainable.
Although the two formulations look quite different, both make guarantees limiting how one component of a system can influence---i.e., interfere with---another.
In authorization logics, this takes the form ``Alice's beliefs can only impact the provability of Bob's beliefs if Bob trusts Alice.''
In information-flow systems---which are mostly defined over programs---changing the value of an input variable~$x$ can only change the value of an output variable~$y$ when the label of~$x$ flows to the label of~$y$.

Both of these notions of non-interference are important.
Consider again the example where Bob's friend list is private but Alice attempts to view his photo.
Because Bob's friend list is private, changing the list should not affect Alice's beliefs.
For instance, Alice should not be affected by Bob adding or removing Cathy.
To enforce this, whether or not Cathy is Bob's friend must not affect the set of Bob's beliefs that Alice \emph{may} learn.
This requires authorization-logic non-interference, since Bob's beliefs should not affect Alice's beliefs unless they communicate.
It also, however, requires information-flow non-interference, since the privacy of Bob's belief is why he is unwilling to communicate.

Gluing together both ideas of non-interference requires understanding the connection between their notions of trust.
As we have discussed, this connection is difficult to formulate, making the non-interference combination harder still.

Our goal in this work is to provide a logic that supports reasoning about both information flow and authorization policies by combining their models of trust to obtain the advantages of both.
To this end, we present the \emph{Flow-Limited Authorization First-Order Logic} (FLAFOL), which
\begin{itemize}
\item provides a notion of trust between principals that can vary depending on information-flow labels,
\item clearly denotes points where communication occurs,
\item uses $\says$ formulae to reason about principals' beliefs, including their beliefs about others' beliefs,
\item is expressive enough to encode real-world authorization policies, and
\item provides a strong security guarantee which combines both authorization-logic and information-flow non-interference.
\end{itemize}

We additionally aim to clarify the foundations of flow-limited authorization (introduced by Arden et al.~\cite{flam}).
We therefore strive to keep FLAFOL's model of principals, labels, and communication as simple as possible.
For example, unlike previous work, we do not require that labels form a lattice.

A final contribution is an implementation of FLAFOL in Coq~\cite{Coq:manual} and formal proofs of all theorems in this paper.
\footnote{The Coq code is available at \url{https://github.com/FLAFOL/flafol-coq}.}
Together these consists of 18,384 lines of Coq code.
For more details, see Appendix~\ref{sec:coq-details}.

We are, of course, not the first to recognize the important interaction of information-flow policies with authorization,
but all prior work in this area is missing at least one important feature.
The three projects that have done the most to combine authorization and information flow are FLAM~\cite{flam}, SecPAL\textsuperscript{+}~\cite{SecPAL,secpal-plus}, and \textsc{Aura}~\cite{aura,aura-if}.
FLAM models trust using information flow, \textsc{Aura} uses DCC~\cite{ccd99,abadi06}, a propositional authorization logic, and SecPAL\textsuperscript{+} places information flow labels on principal-based trust policies, but does not attempt to reason about the combination at all.
Neither FLAM nor SecPAL\textsuperscript{+} can reason about nested beliefs, and both are significantly restricted in what logical forms are allowed.
Finally, FLAM's security guarantees are non-standard and difficult to compare to other languages (see Section~\ref{sec:related}), while \textsc{Aura} relies on DCC's non-interference guarantee which does not apply on any trust relationships outside of those assumed in the static lattice.

The rest of this paper is organized as follows:
In Section~\ref{sec:flafol-example} we discuss three running examples.
This also serves as an intuitive introduction to FLAFOL.
In Section~\ref{sec:system-model} we show how FLAFOL's parameterization allows it to model real systems.
In Section~\ref{sec:proof-system} we detail the FLAFOL proof rules.
In Section~\ref{sec:proof-theory} we discuss the proof theory of FLAFOL, proving important meta-level theorems, including consistency and cut elimination.
In Section~\ref{sec:non-interference} we provide FLAFOL's non-interference theorem.
We discuss related work in Section~\ref{sec:related}, and finally we conclude in Section~\ref{sec:conclusion}.


%% file: by_example.tex
\section{FLAFOL By Example}
\label{sec:flafol-example}

We now examine several examples of authorization policies and how FLAFOL expresses them.
This will serve as a gentle introduction to the main ideas of FLAFOL, and introduce notation and running examples we use throughout the paper.

We explore three main examples in this section:
\begin{enumerate}
\item Viewing pictures on social media
\item Sanitizing data inputs to prevent SQL injection attacks
\item Providing a hospital bill in the presence of reinsurance
\end{enumerate}

Each setting has different requirements, such as defining the meaning of labels in its own way.
The ability of FLAFOL to adapt to each demonstrates its expressive power.
In a new setting, it is often convenient---even necessary---to define constants, functions, and relations beyond those baked into FLAFOL.
FLAFOL supports this by being parameterized over such definitions and having a security guarantee which holds for any parameterization.
We use such symbols freely in our examples to express our intent clearly.
Formally, FLAFOL interprets them using standard proof-theoretic techniques, as we see in Section~\ref{sec:system-model}.

Notably, FLAFOL does not allow computation on terms, so the meaning of functions and constants are axiomatized via FLAFOL formulae.
This allows principals to disagree on how functions behave,
which can be useful in modeling situations where each principal has their own view of some piece of data.

\subsection{Viewing Pictures on Social Media}
\label{sec:looking-at-pictures}

We begin by reconsidering in more detail the example from Section~\ref{sec:introduction} where Alice requests to view Bob's picture on a social-media service.
This service allows Bob to set privacy policies, and Bob made his pictures visible only to his friends.
When Alice makes her request, the service can check if she is authorized by scanning Bob's friend list.
If she is on the list and the photo is available, it shows her the photo.
If she is \emph{not} on Bob's friend list, it shows her \mbox{HTTP 403: Forbidden}.

Bob may choose who belongs in the role of ``friend.''
Following the lead of other authorization logics, FLAFOL represents Bob believing that Alice is his friend as \mbox{$\Bob \says \isFriend(\Alice)$}.
Since $\says$ statements can encompass any formula, we can express the fact that Bob believes that Alice is \emph{not} his friend as $\Bob \says \lnot \isFriend(\Alice)$.

We interpret these statements as Bob's \emph{beliefs}.
This reflects the fact that Bob could be wrong, in the sense that he may affirm formulae with provable negations.
There is no requirement that Bob believes all true things nor that Bob only believe true things (see Section~\ref{sec:proof-system}), so holding an incorrect belief does not require Bob to believe $\False$.
Note that because $\False$ allows us to prove anything, a principal who \emph{does} believe $\False$ will affirm every statement.

Now imagine that, as in Section~\ref{sec:introduction}, the social-media service allows Bob to set a privacy policy on his friend list as well.
As before, Bob can restrict his friend list so that only his friends may learn its contents.
In order to discuss such a policy in FLAFOL, we need a way to express that Bob's friend list is private.
Since, formally, his friend list is a series of beliefs about who his friends are, we must express the privacy of those beliefs.
We view this as giving each belief a \emph{label} describing Bob's policy about who may learn that belief.
Syntactically, we attach this label to the $\says$ connective.
For example, Bob may use the label $\frnd$ to represent the information-security policy ``I will share this with only my friends.''

If he attaches this policy to the beliefs representing his friend list, there is no way to securely prove either \mbox{$\Bob \says*{\ell} \isFriend(\Alice)$} or \mbox{$\Bob \says*{\ell} \lnot\isFriend(\Alice)$} when $\ell$ is less restrictive than $\frnd$.
To see why, imagine what happens when Alice makes her request.
If she is on Bob's friend list, she may again see the photo.
However, if she is not, showing her an HTTP~403 page would leak Bob's private information; Alice would learn that she is not on Bob's friend list, something Bob only shared with his friends.
Since FLAFOL's security guarantee (Theorem~\ref{thm:NI}) shows that every FLAFOL proof is secure, neither option is provable in FLAFOL.
Clearly Bob needs to define a more permissive policy on his friend list.

If Bob's friend list were public, simply checking the list would be enough to prove either of the above statements.
FLAFOL can easily express this by labeling each of Bob's beliefs about $\isFriend$ as $\pub$.  
Another, more subtle, change would be to say that every principal can find out whether \emph{they} are on Bob's friend list, but only Bob's friends can see the rest of the list.
FLAFOL can also express this policy and prove it decidable, but doing so will require significant infrastructure using the technology we will build in Sections~\ref{sec:system-model} and~\ref{sec:proof-system}.
We show how to express this policy in Appendix~\ref{sec:view-pict-soci}.

This example demonstrates how naively reasoning about authorization with information flow can cause leaks, and how FLAFOL can help reason about those beliefs, leading to enforceable policies that 
capture the intent of system developers.

\subsection{Integrity Tracking to Prevent SQL Injection}
\label{sec:prev-sql-inject}

For our second example, imagine a stateful web application.
It takes requests, updates its database, and returns web pages.
In order to avoid SQL injection attacks, the system will only update its database based on high-integrity input.
However, it marks all web request inputs as low integrity, representing the fact that they may contain attacks.
The server has a sanitization function $\san$ that will neutralize attacks, so when it encounters a low-integrity input, it is willing to sanitize that input and endorse the result.

FLAFOL's support for arbitrary implications allows it to easily encode such endorsements.
Let the predicate $\Input(x)$ mean that a value~$x$---possibly taken from a web request---is a database input.
When a user makes a request with database input~$x$, we can thus represent it as $\sys \says*{\untrusted} \Input(x)$.
Here $\untrusted$ represents low-integrity beliefs.
We represent the system's willingness to endorse any sanitized input as:
$$ \sys \says*{\untrusted} \Input(x) \impl \sys \says*{\trusted} \Input(\san(x))$$

This example shows the power of arbitrary implications for expressing authorization and information-flow policies.
It also, however, demonstrates their dangers, since unconstrained downgrades can allow information to flow in unintended ways.
In Section~\ref{sec:non-interference} we will discuss how non-interference (Theorem~\ref{thm:NI}) adapts to these downgrades by weakening its guarantees.

\subsection{Hospital Bills Calculation and Reinsurance}
\label{sec:prov-hosp-bill}

Imagine now that Alice finds herself in the hospital.
Luckily her employer provides health insurance, but they have just switched companies.
Now she has two unexpired insurance cards, and she cannot figure out which one is valid.
Thus, either of two insurers, $I_1$ and $I_2$, may be paying.

Imagine further that Bob's job is to create a correct hospital bill for Alice.
He uses the label $\ell_H$ to determine both who may learn the contents of Alice's bill and who may help determine them.
That is, $\ell_H$ expresses both a confidentiality policy and an integrity policy.
Bob believes that Alice's insurer may help determine the contents of Alice's bill, since they can decide what they are willing to pay for Alice's surgery.

Bob knows that $I_2$ has a reinsurance contract with $I_1$.
This means that if Alice is insured with $I_2$ and the surgery is very expensive, $I_1$ will pay some of the bill.
Thus, $I_1$ may help determine the contents of Alice's hospital bill, even if $I_2$ turns out to be her current insurer.

Bob is willing to accept Alice's insurance cards as evidence that she is insured by either $I_1$ or $I_2$, which we can express as $\Bob \says*{\ell_H} (\canw(I_1, \ell_H) \lor \canw(I_2, \ell_H))$.
Because Bob knows about $I_2$'s reinsurance contract with $I_1$, he knows that if $I_2$ helps determine the contents of Alice's bill, they will delegate some of their power to $I_1$, which we express as $\Bob \says*{\ell_H} (I_2 \says*{\ell_H} \canw(I_1, \ell_H))$.

Bob's beliefs allow him to prove that $I_1$ may help determine the contents of Alice's bill, since by assuming the previous two statements we can prove that $\Bob \says*{\ell_H} \canw(I_1, \ell_H)$.
There are two possible cases: if Bob already believes that $I_1$ can help determine the contents of Alice's bill, we are done.
Otherwise, Bob believes that $I_2$ can help determine the contents of Alice's bill, and so Bob is willing to let $I_2$ delegate their power.
Since he knows that they will delegate their power to $I_1$, he knows that $I_1$ can help determine the contents of Alice's bill in this case as well.
This covers all of the cases, so we can conclude that $\Bob \says*{\ell_H} \canw(I_1, \ell_H)$.

We think of Bob as performing this proof, since it is entirely about Bob's beliefs.
From this point of view, Bob's ability to reason about $I_2$'s beliefs appears to be Bob \emph{simulating} $I_2$.
This ability of one principal to simulate another provides the key intuition to understand the \emph{generalized principal}, a fundamental construct in the formal presentation of FLAFOL (see Section~\ref{sec:system-model}).

We also note that Bob used $I_2$'s beliefs in this proof, even though he does not necessarily trust $I_2$.
However, he \emph{might} trust it if it turns out to be Alice's insurer.
Because Bob trusts $I_2$ in part of the proof but not in general, we refer to this as \emph{discoverable trust}.
FLAFOL's ability to handle discoverable trust makes reasoning about its security properties more difficult, as we see in Section~\ref{sec:non-interference}.

This example shows how disjunctions can be used to express policies when principals do not know the state of the world.
It also demonstrates how disjunctions make it difficult to know how information can flow at any point in time, since we may discover new statements of trust under one branch of a disjunction.
FLAFOL's non-interference theorem adapts to this by considering all declarations of trust that could possibly be discovered in a given context.

\subsection{Further Adapting FLAFOL}
\label{sec:furth-adapt-flaf}

All of the above examples use information-flow labels to express confidentiality policies, integrity policies, or both.
While confidentiality and integrity are mainstay features of information flow tracking, information-flow labels can also express other properties.
For instance, MixT~\cite{mixt} describes how to use information-flow labels to create safe transactions across databases with different consistency models, and the work of Zheng and Myers~\cite{zm05} uses information-flow labels to provide availability guarantees.
FLAFOL allows such alternative interpretations of labels by using an abstract \emph{permission model} to give meaning to labels.

By default, the permissions gain meaning only through their behavior in context, but they are able to encode and reason about a wide variety of authorization mechanisms.
In Section~\ref{sec:system-model}, we see how FLAFOL can be used to reason about capabilities, and in Appendix~\ref{sec:exampl-perm-models} we discuss a model closer to military classification.


%% file: using_flafol.tex
\section{Using FLAFOL}
\label{sec:system-model}

In this section, we examine how to use FLAFOL to reason about real systems.
To do this, we look at a fictional verified-distributed-systems designer Dana.
She wants to formally prove that confused-deputy attacks are impossible in her capability-based system with copyable, delegatable read capabilities.
Dana employs a six-step process to reason about her system in FLAFOL:
\begin{enumerate}[leftmargin=*]
  \item Decide on a set $\mathcal{S}$ of \emph{sorts} of data she wants to represent.
  \item Choose a set $\mathcal{F}$ of \emph{function symbols} representing operations in the system, and give those operations types.
  \item Choose a set $\mathcal{R}$ of \emph{relation symbols} representing atomic facts to reason about, and give the relations types.
  \item Develop axioms that encode meaning for these relationships.
  \item Specify meta-level theorems stating her desired properties.
  \item Prove that those meta-level theorems hold.
\end{enumerate}

\paragraph{Sorts} First, Dana decides on what sorts of data she wants to represent.
We can think of \emph{sort} as the logic word for ``type.''
FLAFOL is defined with respect to a set $\mathcal{S}$ of sorts that must include at least \labl{} and \prin{}, but may contain more.
Dana wants to reason about capability tokens that grant read access to data, so she also includes a sort named \tkn{}.

Dana uses the \prin{} sort to represent system principals, but conceptually divides the \labl{} sort into \conf{} and \integ{}, two sorts which she also adds.
Each \conf{} value defines a confidentiality policy which may be applied to many pieces of data.
A capability (which is always public itself) grants read access to data governed by one or more such policies.
She uses the \integ{} sort to represent integrity policies on tokens themselves.
We will see below how she can enforce \mbox{$\labl = \conf \times \integ$}.

\paragraph{Function Symbols}
Dana next decides on operations she wants to reason about.
This is also her chance to define constants using nullary operations.
Formally, FLAFOL is defined with respect to an arbitrary set $\mathcal{F}$ of \emph{function symbols}.
Each function comes equipped with a \emph{signature}, or type, expressing when it can be applied.

Dana considers what information she needs about a given token.
She needs a way to determine which confidentiality level a token grants permission to read, the integrity of that token, and which principal is the token's \emph{root of authority}---that is, who created the token.
She thus creates three function symbols:
\begin{align*}
  \tknconf & : \tkn \to \conf \\
  \tkninteg & : \tkn \to \integ \\
  \rootofauth & : \tkn \to \prin
\end{align*}
She also needs to be able to determine the integrity that a principal commands, so she includes a function symbol \mbox{$\integof : \prin \to \integ$}.
Finally, since a token can potentially be transferred to anyone in her system, she creates a constant \mbox{$\pub : \conf$} to represent this.

Dana wants to enforce that labels are pairs of confidentiality and integrity.
She therefore creates two ``projection'' function symbols $\projconf$ and $\projinteg$, and a third pair symbol $(\_,\_)$ with the following signatures:
\begin{align*}
  \projconf & : \labl \to \conf \\
  \projinteg & : \labl \to \integ \\
  (\_,\_) & : \conf \to \integ \to \labl
\end{align*}
The first two ensure that labels contain a confidentiality and an integrity, while pairing allows creation of labels from a confidentiality with an integrity.
This makes labels pairs of confidentiality and integrity.
Dana also adds axioms corresponding to the $\eta$ and $\beta$ laws for pairs.

\paragraph{Relation Symbols} Dana can now choose relations representing facts that she wants to reason about.
Along with sorts and functions, FLAFOL is defined with respect to a set $\mathcal{R}$ of \emph{relation symbols}, allowing it to reason about more facts.
The set $\mathcal{R}$ must include at least flows-to~($\flowsto$), $\canr$, and $\canw$, but may contain more.
We call these required relations \emph{permissions} because they define the trust relationships governing communication.
The relation~$\ell \flowsto \ell'$ means information with label~$\ell$ can affect information with label~$\ell'$, $\canr(p, \ell)$ means that principal~$p$ may learn beliefs with label~$\ell$, and $\canw(p, \ell)$ means~$p$ may influence beliefs with label~$\ell$.

Dana is able to use these relations to define the permissions her capability tokens grant.
She also includes a fourth relation in $\mathcal{R}$, $\hastkn(\prin, \tkn)$, defining token possession: if $\hastkn(p, t)$, then principal $p$ has (a copy of) token~$t$.

\paragraph{Axioms}
Dana describes the behavior of her system with axioms that use the sorts, functions, and relations she defined above.
These should be \emph{consistent}, in the sense that they do not allow a derivation of \False{}.
Theorem~\ref{thm:pos-consistent} in Section~\ref{sec:consistency} gives conditions under which all of the axioms that we will discuss in this section are consistent.

Dana uses three main axioms: one describing how tokens may be copied and delegated, one describing when one principal may read another's beliefs, and one describing when a principal may affect another's beliefs.
She may use more axioms if she likes---e.g., to capture principals' beliefs about permitted flows between labels.

Dana's first axiom allows any principal to copy any capability it holds and give that copy to another principal:
$$\begin{array}{l}
  \forallexp{q}{\prin}{\forallexp{t}{\tkn}{}} \\
  \quad\left(\begin{array}{c@{}l}
    \existsexp{p}{\prin}{} & \hastkn(p, t) \\
    & \mathrel{\land} p \says*{(\pub, \tkninteg(t))} \hastkn(q, t)
  \end{array}\right) \impl \hastkn(q, t)%
\end{array}$$
This says that, for principals $p$ and $q$, if $p$ holds a read capability token $t$, $p$ can pass $t$ to $q$.
To do so, $p$ must affirm that $q$ has $t$ at a public label with the integrity of the token.
Note that the use of $\pub$ here means Dana's system must allow everyone to learn whenever one principal copies a token and passes it to another.

Dana's second axiom defines when a principal $p$ allows $q$ to read a belief of $p$'s labeled $\ell$.
First, $p$ checks that $q$ has a token, and that $p$ believes that the token gives read access to something at least as confidential as $\ell$.
Second, $p$ checks to make sure that the token's root authority may influence this belief:
$$\begin{array}{l}
  \forallexp{q}{\prin}{\forallexp{\ell}{\labl}{\forallexp{p}{\prin}{\forallexp{\ell'}{\labl}{}}}} \\
  \quad\left(
    \begin{array}{c@{}l}
      \existsexp{t}{\tkn}{} & \hastkn(q,t) \\
      & \mathrel{\land} p \says*{\ell'} \projconf(\ell) \flowsto \tknconf(t) \\
      & \mathrel{\land} p \says*{\ell'} \canw(\rootofauth(t), \ell')
    \end{array}\right) \impl p \says*{\ell'}(\canr(q,\ell))
  \end{array}$$
More formally, it says that if~$q$ holds some token~$t$ and $p$~believes both that~$t$ grants read permissions for~$\ell$'s confidentiality and that the root of authority for~$t$ can influence~$p$'s beliefs at~$\ell'$, then $p$ will allow $q$ to read $\ell$.
This defines what it means for a principal ($p$ here) to believe that a token grants read access to their data.
Dana now needs to make sure that whenever a read access is granted in her system, not only does the principal who gets read access have a token, but that the principal who owns the data does indeed believe that the token grants read access to that data.  

Finally, her third axiom states that one principal~$p$ believes that another principal~$q$, can write a label~$\ell$ if $p$ believes that the integrity of $q$ flows to the integrity of $\ell$:
$$\begin{array}{l}
  \forallexp{q}{\prin}{\forallexp{\ell}{\labl}{\forallexp{p}{\prin}{\forallexp{\ell'}{\labl}{}}}} \\
  \quad p \says*{\ell'} \left(\integof(q) \flowsto \projinteg(\ell)\right) \impl p \says*{\ell'} \left(\canw(q, \ell)\right)
\end{array}$$
Dana then needs to make sure that write accesses are only granted to principals with high enough integrity.  

\paragraph{Metatheoretic Properties}
Dana has now created a model of her system, so she can use it to state and prove properties of her system as meta-theorems.
Luckily, Rajani, Garg, and Rezk~\cite{RajaniGargRezk16} have shown that information-flow integrity tracking with a non-interference result is sufficient to avoid confused deputy attacks with capability systems.
Therefore Theorem~\ref{thm:NI} provides the guarantees she needs.

\begin{figure}
  \centering
  \small
  $$\begin{syntax}
    \synclass{Sorts}{\sigma}{\labl \alt \prin \alt \dotsb}
    \emptysynclass{Labels}{\ell}
    \emptysynclass{Principals}{p, q, r}

    \synclass{Functions}{f}{\dotsb}
    \synclass{Relations}{R}{\canr(\prin, \labl)}
    \altclause{\canw(\prin, \labl)}
    \altclause{\labl \flowsto \labl \alt \dotsb}
    \synclass{$\sigma$-terms}{t}{x \alt f(t_1, \ldots, t_n)}
    \synclass{Formulae}{\varphi, \psi, \chi}{R(t_1, \dotsc, t_n)}
    \altclause{\True \alt \False}
    \altclause{\varphi \land \psi \alt \varphi \lor \psi \alt \varphi \mimpl* \psi}
    \altclause{\forallexp*{\varphi} \alt \existsexp*{\varphi}}
    \altclause{p \says*{\ell} \varphi}

    \synclass{\parbox[c]{5em}{\centering Generalized Principals}}{g}{\modalG{\ell} \alt g \cdot \modal{p}{\ell}}
  \end{syntax}$$
  \caption{FLAFOL Syntax}
  \label{fig:syntax}
\end{figure}

\paragraph{FLAFOL Syntax} This example demonstrates FLAFOL's flexibility as a powerful tool for reasoning about authorization mechanisms in the presence of information-flow policies.
We saw that, since FLAFOL is defined with respect to the three sets~$\mathcal{S}$,~$\mathcal{F}$, and~$\mathcal{R}$, it can express the key components of a system.
This parameterized definition gives rise to the formal FLAFOL syntax in Figure~\ref{fig:syntax}.

In order to use the function and relation symbols and incorporate axioms, FLAFOL allows proofs to occur in a context.
FLAFOL additionally includes rules requiring flows-to to be reflexive and transitive, placing a preorder on the \labl{} sort,
\footnote{Many information-flow tools require their labels to form a lattice.
We find that a preorder is sufficient for FLAFOL's design and guarantees, so we decline to impose additional structure.
In Section~\ref{sec:label-lattice-details} we show that enforcing a lattice structure is both simple and logically consistent.}
and requiring $\canr$ and $\canw$ to respect a form of variance.
If $\ell_1 \flowsto \ell_2$ and Alice can read data $A$ with label $\ell_2$, then she may learn information about data with label $\ell_1$ used to calculate $A$.
This means she should also be able to read data with label $\ell_1$.
Thus, $\canr$ must (contravariantly) respect the preorder on labels.
Similarly, if Alice can help determine some piece of data $B$ labeled with $\ell_1$, she can influence any data labeled with $\ell_2$ that is calculated from $B$, so Alice should be able to help determine data labeled at $\ell_2$.
Thus, $\canw$ must (covariantly) respect the preorder on labels.

Figure~\ref{fig:flows-to-permission} presents these rules formally.
We give the proof rules in the form of a sequent calculus.
The trailing $\atl g$ represents \emph{who} affirms that formula in the proof, similarly to how $\says$ formulae represent who affirms a statement at the object level.
Unlike $\says$ formulae, these meta-level objects---which we call \emph{generalized principals}---encode arbitrary reasoners, including possibly-simulated principals.

Recall from Section~\ref{sec:prov-hosp-bill} that we can think of some proofs as being performed by principals if those proofs entirely involve that principal's beliefs.
In that example, Bob reasoned about his belief that another principal, the insurer $I_2$, trusted a third principal, the insurer $I_3$.
We think of this ability to reason about the beliefs of others as the ability to \emph{simulate} other principals.
In fact, because principals' beliefs are segmented by labels, principals can have multiple simulations of the same other principal.

This suggests that FLAFOL captures the reasoning of principals \emph{at some level of simulation}.
A generalized principal is a stack of principal/label pairs, representing a stack of simulators and simulations.
The empty stack, written $\modalG{}$, represents \emph{ground truth}.
A stack with one more level, written $g \cdot \modal{p}{\ell}$, represents the beliefs of $p$ at level $\ell$ according to the generalized principal $g$.
Figure~\ref{fig:syntax} contains the formal grammar for generalized principals.

\begin{figure}
  \small
  \begin{mathpar}
    \infer*[left=FlowsToRefl]{ }{\Gamma \proves \belief{\ell \flowsto \ell}{g}} \and
    \infer*[left=FlowsToTrans]{\Gamma \proves \belief{\ell_1 \flowsto \ell_2}{g} \\ \Gamma \proves \belief{\ell_2 \flowsto \ell_3}{g}}{\Gamma \proves \belief{\ell_1 \flowsto \ell_3}{g}} \and
    \infer*[left=CRVar]{\Gamma \proves \belief{\canr(p, \ell_2)}{g}\\ \Gamma \proves \belief{\ell_1 \flowsto \ell_2}{g}}{\Gamma \proves \belief{\canr(p, \ell_1)}{g}} \and
    \infer*[left=CWVar]{\Gamma \proves \belief{\canw(p, \ell_1)}{g}\\ \Gamma \proves \belief{\ell_1 \flowsto \ell_2}{g}}{\Gamma \proves \belief{\canw(p, \ell_2)}{g}}
  \end{mathpar}
  \caption{Permission Rules}
  \label{fig:flows-to-permission}
\end{figure}


%% file: proof_system.tex
\section{Proof System}
\label{sec:proof-system}

So far, we have discussed the intuitions behind FLAFOL and its syntax.
Here we introduce FLAFOL formally.
Unfortunately, we cannot examine every aspect of FLAFOL's formal presentation in detail, though interested readers should see Appendix~\ref{sec:full-flafol-proof}.
Instead, we discuss the most novel and most security-relevant aspects of FLAFOL's design.

FLAFOL sequents are of the form $\Gamma \proves \belief{\varphi}{g}$, where $\Gamma$ is a context containing beliefs.
This means that the FLAFOL proof system manipulates beliefs, as described in Section~\ref{sec:system-model}.
Readers familiar with sequent calculus may recognize that FLAFOL is intuitionistic, as there is only one belief on the right side of the turnstile.
\footnote{Recall that we argued in Section~\ref{sec:looking-at-pictures} that reasoning about authorization and information-flow security together is naturally intuitionistic, since we cannot securely conclude $\varphi$ or $\lnot\varphi$ in some naturally-occurring contexts.}

Sequent calculus rules tend to manipulate beliefs either on the left or the right side of the turnstile.
For instance, consider the FLAFOL rules for disjunctions:
\begin{mathpar}
    \infer*[left=OrL]{
      \Gamma, \belief{\varphi}{g} \proves \belief{\chi}{g'} \\
      \Gamma, \belief{\psi}{g} \proves \belief{\chi}{g'}}
    {\Gamma, (\belief{\varphi \lor \psi}{g}) \proves \belief{\chi}{g'}} \and
    \infer*[left=OrR1]{\Gamma \proves \belief{\varphi}{g}}{\Gamma \proves \belief{\varphi \lor \psi}{g}} \and
     \infer*[left=OrR2]{\Gamma \proves \belief{\psi}{g}}{\Gamma \proves \belief{\varphi \lor \psi}{g}}
\end{mathpar}
We find it easiest to read left rules ``up'' and right rules ``down.''
With this reading, the \textsc{OrL} rule tells us how to use an assumption of the form $\belief{\varphi \lor \psi}{g}$ in order to prove a belief~$\belief{\chi}{g'}$ by performing case analysis.
That is, \textsc{OrL} tells us how to prove $\belief{\chi}{g'}$ assuming $\belief{\varphi \lor \psi}{g}$ if we can prove that $\belief{\chi}{g'}$ assuming $\belief{\varphi}{g}$ and separately assuming $\belief{\psi}{g}$.

The \textsc{OrR1}~rule takes a proof of~$\belief{\varphi}{g}$ and uses it to prove~$\belief{\varphi \lor \psi}{g}$.
The \textsc{OrR2}~rule is symmetric, so it takes a proof of~$\belief{\psi}{g}$ and uses it to prove~$\belief{\varphi \lor \psi}{g}$.
\footnote{For readers interested in learning more about sequent calculus, we recommend MIT's interactive tool for teaching sequent calculus as a tutorial~\cite{MITSequentCaluclus}.}

Note that these rules (along with the $\says$ rules discussed below) allow $\says$ to distribute over disjunctions.
That is, given $p \says*{\ell} (\varphi \lor \psi)$, we can prove $(p \says*{\ell} \varphi) \lor (p \says*{\ell} \psi)$.
In an intuitionistic logic like FLAFOL, disjunctions must be a proof of one side or the other.
The proof that $\says$ distributes over $\lor$ then says that if $p$ has evidence of either $\varphi$ or $\psi$, then $p$ can examine this evidence to discover whether it is evidence of $\varphi$ or of~$\psi$.

\begin{figure*}
  \small
  \begin{mathpar}
    \infer*[left=FalseL]{ }{\Gamma, \belief{\False}{g} \proves \belief{\varphi}{g \cdot g'}}
    \\
    \infer*[left=ImpL]{
      \Gamma \proves \belief{\varphi}{\modalG{\ell}} \\
      \Gamma, \belief{\psi}{g} \proves \belief{\chi}{g'} \\
    }{\Gamma, (\belief{\varphi \mimpl* \psi}{g}) \proves \belief{\chi}{g'}} \and
    \infer*[left=ImpR]{\Gamma, \belief{\varphi}{\modalG{\ell}} \proves \belief{\psi}{g}}{\Gamma \proves \belief{\varphi \mimpl* \psi}{g}}
    \\
    \infer*[left=SaysL]{\Gamma, \belief{\varphi}{g \cdot \modal{p}{\ell}} \proves \belief{\psi}{g'}}{\Gamma, \belief{p \says*{\ell} \varphi}{g} \proves \belief{\psi}{g'}} \and
    \infer*[left=SaysR]{\Gamma \proves \belief{\varphi}{g \cdot \modal{p}{\ell}}}{\Gamma \proves \belief{p \says*{\ell} \varphi}{g}}
    \\
    \infer*[left=VarR]{
      \Gamma \proves \belief{\varphi}{g \cdot \modal{p}{\ell'} \cdot g'} \\\\
      \Gamma \proves \belief{\ell' \flowsto \ell}{g \cdot \modal{p}{\ell}}}
    {\Gamma \proves \belief{\varphi}{g \cdot \modal{p}{\ell} \cdot g'}} \and
    \infer*[left=FwdR]{
      \Gamma \proves \belief{\varphi}{g \cdot \modal{p}{\ell} \cdot g'}\\\\
      \Gamma \proves \belief{\canr(q, \ell)}{g \cdot \modal{p}{\ell}}\\
      \Gamma \proves \belief{\canw(p, \ell)}{g \cdot \modal{q}{\ell}}}
    {\Gamma \proves \belief{\varphi}{g \cdot \modal{q}{\ell} \cdot g'}} \and
  \end{mathpar}
  \caption{Selected FLAFOL Proof Rules}
  \label{fig:selected-proof-rules}
\end{figure*}

Most of the rules of FLAFOL are standard rules for first-order logic, but with generalized principals included to indicate who believes each formula.
For instance, the rules for disjunctions above were likely familiar to those who know sequent calculus.

Figure~\ref{fig:selected-proof-rules} contains FLAFOL rules selected for discussion.
The first, \textsc{FalseL}, tells us how to use $\False$ as an assumption.
In standard intuitionistic first-order logic, this is simply the principle of Ex Falso: if we assume $\False$, we can prove anything.
In FLAFOL, a generalized principal who assumes false is willing to affirm any formula.
This includes statements about other principals, so \textsc{FalseL} extends the generalized principal arbitrarily.
We use $g \cdot g'$ as notation for extending the generalized principal $g$ with a list of principal-label pairs, denoted $g'$.

The implication rules \textsc{ImpR} and \textsc{ImpL} interpret the premise of an implication as ground truth, while the generalized principal who believes the implication believes the consequent.
In particular, this means that $\says$ statements do not distribute over implication as one might expect, i.e., $p \says*{\ell} (\varphi \mimpl* \psi)$ does not imply that $(p \says*{\ell} \varphi) \mimpl* (p \says*{\ell} \psi)$.
Instead, \mbox{$p \says*{\ell} (\varphi \mimpl* \psi)$} implies \mbox{$\varphi \mimpl* (p \says*{\ell} \psi)$}.
We can thus think of implications as \emph{conditional} knowledge.
That is, if a generalized principal $g$ believes~$\varphi \mimpl* \psi$, then~$g$ believes $\psi$ conditional on $\varphi$ being true about the system.

We can still form implications about generalized principals' beliefs, but we must insert appropriate $\says$ statements into the premise to do so.
In Section~\ref{sec:impl-comm-cut-elim}, we discuss how this semantics is necessary for both our proof theoretic and our security results.

The next two rules of Figure~\ref{fig:selected-proof-rules}, \textsc{SaysR} and \textsc{SaysL}, are the only rules which specifically manipulate $\says$ formulae.
Essentially, generalized principals allow us to delete the $\says$ part of a formula while not forgetting who said it.
Thus, generalized principals allow us to define sequent calculus rules once for every possible reasoner.

The final rules, \textsc{VarR} and \textsc{FwdR}, define communication in FLAFOL.
Both manipulate beliefs on the right and have corresponding left rules, which act contravariantly and can be found in Appendix~\ref{sec:full-flafol-proof}.

Information-flow communication is provided by the variance rule \textsc{VarR}.
This can be thought of like the variance rules used in subtyping.
Most systems with information-flow labels do not have explicit variance rules, but instead manipulate relevant labels in every rule.
By adding an explicit variance rule, we not only simplify every other FLAFOL rule, we also remove the need for the label join and meet operators that are usually used to perform the label manipulations.
Others have noted that adding explicit variance rules simplifies the design of the rest of the system~\cite{vsi96,DCC-perspective}, but it remains an unusual choice.

The forwarding rule \textsc{FwdR} provides authorization-logic-style communication.
In FLAFOL, $p$ can forward a belief at label $\ell$ to $q$ if:
\begin{itemize}
\item $p$ is willing to send its beliefs at label $\ell$ to $q$, denoted $p \says*{\ell} \canr(q, \ell)$, and
\item $q$ is willing to allow $p$ to determine its beliefs at label $\ell$, denoted $q \says*{\ell} \canw(p, \ell)$.
\end{itemize}
After establishing this trust, $p$ can package up its belief and send it to $q$, who will believe it at the same label.


%% file: proof_theory.tex
\section{Proof Theory}
\label{sec:proof-theory}

In this section, we evaluate FLAFOL's logical design.
We show that FLAFOL has the standard sequent calculus properties of (positive) consistency and cut elimination
and discuss fundamental limitations that inform our unusual implication semantics.
We also develop a new proof-theoretic tool, \emph{compatible supercontexts}, for use in our non-interference theorem in Section~\ref{sec:non-interference}.

\subsection{Consistency}
\label{sec:consistency}

One of the most important properties about a logic is consistency, meaning it is impossible to prove $\False$.
This is not possible in an arbitrary context, since one could always assume $\False$.
One standard solution is to limit the theorem to the empty context.
By examining the FLAFOL proof rules, however, we see that it is only possible to prove $\False$ by assumption or by Ex Falso.
Either method requires that $\False$ already be on the left-hand side of the turnstile, so if $\False$ can never get there, then it should be impossible to prove.

To understand when $\False$ can appear on the left-hand side of the turnstile, we note that formulae on the left tend to stay on the left and formulae on the right tend to stay on the right.
The only exception is the implication rules \textsc{ImpL} and \textsc{ImpR} which move the premise of the implication to the other side.
The fact that no proof rule allows us to change either side of the sequent arbitrarily gives useful structure to proofs.
To handle implications, however, we must keep track of their nesting structure, which we do by considering \emph{signed formulae}.
We call a formula in a sequent \emph{positive} if it appears on the right side of the turnstile and \emph{negative} if it appears on the left.
If~$\varphi$ is positive we write~$\varphi^+$, and if~$\varphi$ is negative we write~$\varphi^-$.

\input{signed-subformla}

This property allows us to prove an important result about the consistency of FLAFOL.

\begin{thm}[Positive Consistency]
  \label{thm:pos-consistent}
  For any context $\Gamma$, if $$\False^- \nleq \varphi^-~\text{for all}~\belief{\varphi}{g} \in \Gamma$$ then $\Gamma \nproves \belief{\False}{g'}$.
\end{thm}

The proof follows by induction on the FLAFOL proof rules.
Note that formulae which do not contain $\False$ as a negative subformula are called \emph{positive} formulae, explaining the name.

We get the result with an empty context as a corollary.
This states that $\False$ is not a theorem of FLAFOL.
\begin{cor}[Consistency] $\nproves \belief{\False}{g}$
\end{cor}

\label{sec:label-lattice-details}
Theorem~\ref{thm:pos-consistent} demonstrates that a variety of useful constructs are logically consistent.
For instance, we can add a lattice structure to FLAFOL's labels.
We can define join ($\sqcup$) and meet ($\sqcap$) as binary function symbols on labels and $\top$ and $\bot$ as label constants.
Then we can simply place the lattice axioms (e.g., $\forallexp{\ell}{\labl}{\ell \flowsto \top}$) in our context to achieve the desired result.
Since none of the lattice axioms include $\False$, Theorem~\ref{thm:pos-consistent} ensures that they are consistent additions to the logic.

\subsection{Simulation}
\label{sec:simulation-1}

\input{simulation}

\subsection{Compatible Supercontexts}
\label{sec:compat-supercontext}

To prove Theorem~\ref{thm:pos-consistent} we needed to consider the possible locations of \emph{formulae} within a sequent, but in Section~\ref{sec:non-interference} we will need to reason about the possible locations of \emph{beliefs}.
To enable this, we introduce the concept of a \emph{compatible supercontext} (CSC).
Informally, the CSCs of a sequent are those contexts that contain all of the information in the current context, along with any counterfactual information that can be considered during a proof.
Intuitively, the rules \textsc{OrL} and \textsc{ImpL} allow a generalized principal to consider such information by using either side of a disjunction or the conclusion of an implication.
If it is possible to consider such a counterfactual, there is a CSC which contains it.
We use the syntax \mbox{$\Delta \ll \Gamma \vdash \belief{\varphi}{g}$} to denote that $\Delta$ is a CSC of the sequent $\Gamma \proves \belief{\varphi}{g}$.
Figure~\ref{fig:sel-compat-superctxt-rules} contains selected rules for CSCs.
The full CSC relation can be found in Appendix~\ref{sec:comp-superc}.

\begin{figure}
  \small
  \begin{mathpar}
    \infer*[left=CSCRefl]{ }{\Gamma \ll  \Gamma \proves  \belief{\varphi}{g}} \and
    \infer*[left=CSCUnion]{\Delta_1 \ll  \Gamma \proves \belief{\varphi}{g}\\ \Delta_2 \ll \Gamma \vdash \belief{\varphi}{g}}{\Delta_1 \cup \Delta_2 \ll \Gamma \vdash \belief{\varphi}{g}} \and
  \end{mathpar}
  \begin{mathpar}
    \infer*[left=CSCOrL1]{\Delta \ll \Gamma, \belief{\varphi}{g} \proves \belief{\chi}{g'}}{\Delta \ll \Gamma, (\belief{\varphi \lor \psi}{g}) \proves \belief{\chi}{g'}} \and
    \infer*[left=CSCImpR]{\Delta \ll \Gamma, \belief{\varphi}{\modalG{}} \proves \belief{\psi}{g}}{\Delta \ll \Gamma \proves \belief{\varphi \mimpl* \psi}{g}} \and
  \end{mathpar}
  \caption{Selected Rules for Compatible Supercontexts}
  \label{fig:sel-compat-superctxt-rules}
\end{figure}

Since all of the information in $\Gamma$ has already been discovered by the generalized principal who believes that information, we require that $\Gamma \ll \Gamma \proves \belief{\varphi}{g}$ with \textsc{CSCRefl}.

If we can discover two sets of information, we can discover everything in the union of those sets using \textsc{CSCUnion}.
This rule feels different from the others, since it axiomatizes certain \emph{properties} of CSCs.
We conjecture that there is an alternative presentation of CSCs where we can prove this rule.

The rest of the rules for CSCs essentially follow the proof rules, so that any belief added to the context during a proof can be added to a CSC.
For instance \textsc{CSCOrL1} and \textsc{CSCOrL2} allow either branch of an assumed disjunction to be added to a CSC, following the two branches of the \textsc{OrL} rule of FLAFOL.

If a context appears in a proof of a sequent, then it is a CSC of that sequent.
We refer to this as the \emph{compatible-supercontext property} (CSC property).
\begin{thm}[CSC Property]
  If \mbox{$\Delta \proves \belief{\psi}{g'}$} appears in a proof of $\Gamma \proves \belief{\varphi}{g}$, then $\Delta \ll \Gamma \proves \belief{\varphi}{g}$.
\end{thm}

\subsection{Cut Elimination}
\label{sec:cut-elimination}

In constructing a proof, it is often useful to create a lemma, prove it separately, and use it in the main proof.
If we both prove and use the lemma in the same context, the main proof follows in that context as well.
We can formalize this via the following rule:
$$\infer*[left=Cut]{\Gamma \vdash \belief{\varphi}{g_1}\\ \Gamma, \belief{\varphi}{g_1} \vdash \belief{\psi}{g_2}}{\Gamma \proves \belief{\psi}{g_2}}$$

This rule is enormously powerful.
It allows us to not only create lemmata to use in a proof, but also simply prove things whose other proofs are complicated and non-obvious.
For instance, consider the rule
$$\infer*[left=UnsaysR]{\Gamma \proves \belief{p \says*{\ell} \varphi}{g}}{\Gamma \proves \belief{\varphi}{g \cdot \modal{p}{\ell}}}$$
We can show that this rule is \emph{admissible}---meaning any sequent provable with this rule is provable without it---
by cutting a proof of the sequent $\Gamma \proves \belief{p \says*{\ell} \varphi}{g}$ with the following proof:
\footnote{Not only can \textsc{UnsaysR} be proven without \textsc{Cut} (as can all FLAFOL proofs), it is actually important for proving cut elimination. See the Coq code.}
$$\infer*[left=SaysL]{
          \infer*[left=Ax]{ }{\belief{\varphi}{g \cdot \modal{p}{\ell}} \proves \belief{\varphi}{g \cdot \modal{p}{\ell}}}}
        {\belief{p \says*{\ell} \varphi}{g} \proves \belief{\varphi}{g \cdot \modal{p}{\ell}}}$$

However, the \textsc{Cut} rule allows an arbitrary formula to appear on both sides of the turnstile in a proof.
That formula may not even be a subformula of anything in the sequent at the root of the proof-tree!
This would seemingly destroy the CSC property that FLAFOL enjoys, and which we rely on in order to prove FLAFOL's security results.
As is standard in sequent calculus proof theory, we show that \textsc{Cut} can be admitted, allowing FLAFOL the proof power of \textsc{Cut} while maintaining the analytic power of the CSC property.

\begin{thm}[Cut Elimination]
  \label{thm:cut-elim}
  The \textsc{Cut} rule is admissible.
\end{thm}

To prove Theorem~\ref{thm:cut-elim}, we first normalize each FLAFOL proof and then induct on the formula $\varphi$ followed by each proof in turn.
Both of these inductions are very involved.
Appendix~\ref{sec:coq-details} contains more details.

This theorem is one of the key theorems of proof theory~\cite{takeuti, proofs-and-types}.
Frank Pfenning has called it ``[t]he central property of sequent calculi''~\cite{structural-cut-elim}.
From the propositions-as-types perspective, cut elimination is preservation of types under substitution.

\subsection{Implications and Communication}
\label{sec:impl-comm-cut-elim}

Recall from Section~\ref{sec:proof-system} how we interpret implication formulae such as \mbox{$\Alice \says*{\ell} (\varphi \mimpl* \psi)$}: if $\varphi$ is true about the system, then Alice believes $\psi$ at label $\ell$.
We can now see why we use this interpretation of implication.
In particular, we consider replacing \textsc{ImpL} and \textsc{ImpR} with the following rules:
\begin{mathpar}
  \infer*[left=$\textsc{ImpL}'$] {\Gamma \proves \belief{\varphi}{g}\\ \Gamma, \belief{\psi}{g} \proves \belief{\chi}{g'}}{\Gamma, \belief{(\varphi \mimpl* \psi)}{g} \proves \belief{\chi}{g'}} \and
  \infer*[left=$\textsc{ImpR}'$] {\Gamma, \belief{\varphi}{g} \proves \belief{\psi}{g}}{\Gamma \proves \belief{\varphi \mimpl* \psi}{g}}
\end{mathpar}
Doing so allows us to prove that $\says$ distributes over implications, as we can see in Figure~\ref{fig:says-over-impl}.
It also allows us to prove that $\says$ \emph{un-distributes} over implication, as we see in Figure~\ref{fig:says-un-over-impl}.
While $\textsc{ImpL}'$, $\textsc{ImpR}'$, and the $\says$ distribution results may all appear sensible, they actually cause security bugs and make cut elimination impossible.

\begin{figure*}
  \begin{minipage}{\textwidth}
    $$
    \infer*[left=SaysL]{
      \infer*[Left=$\textsc{ImpR}'$]{
        \infer*[Left=$\textsc{ImpL}'$\hspace{2.5em},leftskip=2.2em,rightskip=2.2em]{
          \infer*[Left=SaysL]{
            \infer*[Left=Ax]{ }{\belief{\varphi}{g \cdot \modal{p}{\ell}} \proves \belief{\varphi}{g \cdot \modal{p}{\ell}}}
          }{\belief{p \says*{\ell} \varphi}{g} \proves \belief{\varphi}{g \cdot \modal{p}{\ell}}} \\
          \infer*[Right=SaysR]{
            \infer*[Right=Ax]{ }{\belief{\psi}{g \cdot \modal{p}{\ell}} \proves \belief{\psi}{g \cdot \modal{p}{\ell}}}
          }{\belief{\psi}{g \cdot \modal{p}{\ell}} \proves \belief{p \says*{\ell} \psi}{g}}
        }{\belief{(\varphi \mimpl* \psi)}{g \cdot \modal{p}{\ell}}, \belief{p \says*{\ell} \varphi}{g} \proves \belief{p \says*{\ell} \psi}{g}}
      }{\belief{(\varphi \mimpl* \psi)}{g \cdot \modal{p}{\ell}} \proves \belief{(p \says*{\ell} \varphi) \mimpl* (p \says*{\ell} \psi)}{g}}
    }{\belief{p \says*{\ell} (\varphi \mimpl* \psi)}{g} \proves \belief{(p \says*{\ell} \varphi) \mimpl* (p \says*{\ell} \psi)}{g}}
    $$
  \end{minipage}
  \caption[$\says$ over implication]{Proof that $\textsc{ImpL}'$ and $\textsc{ImpR}'$ allow $\says$ to distribute over implication.}
  \label{fig:says-over-impl}
\end{figure*}

\begin{figure*}
  \begin{minipage}{\textwidth}
    $$
    \infer*[left=SaysR]{
      \infer*[Left=$\textsc{ImpR}'$\hspace{1em},leftskip=1em,rightskip=1em]{
        \infer*[Left=$\textsc{ImpL}'$\hspace{1.5em},leftskip=1.5em,rightskip=1.5em]{
          \infer*[Left=SaysR]{
            \infer*[Left=Ax]{ }{\belief{\varphi}{g \cdot \modal{p}{\ell}} \proves \belief{\varphi}{g\cdot\modal{p}{\ell}}}
          }{\belief{\varphi}{g \cdot \modal{p}{\ell}} \proves \belief{p \says*{\ell} \varphi}{g}}\\
          \infer*[Right=SaysL]{
            \infer*[Right=Ax]{ }{\belief{\psi}{g \cdot \modal{p}{\ell}} \proves \belief{\psi}{g \cdot \modal{p}{\ell}}}
          }{\belief{p \says*{\ell} \psi}{g} \proves \belief{\psi}{g \cdot \modal{p}{\ell}}}
        }{\belief{(p \says*{\ell} \varphi) \mimpl* (p \says*{\ell} \psi)}{g}, \belief{\varphi}{g \cdot \modal{p}{\ell} \proves \belief{\psi}{g \cdot \modal{p}{\ell}}}}
      }{\belief{(p \says*{\ell} \varphi) \mimpl* (p \says*{\ell} \psi)}{g} \proves \belief{(\varphi \mimpl* \psi)}{g \cdot \modal{p}{\ell}}}
    }{\belief{(p \says*{\ell} \varphi) \mimpl* (p \says*{\ell} \psi)}{g} \proves \belief{p \says*{\ell}(\varphi \mimpl* \psi)}{g}}
    $$
  \end{minipage}
  \caption[$\says$ over implication, inverted]{Proof that $\textsc{ImpL}'$ and $\textsc{ImpR}'$ allow $\says$ to undistribute over implication.}
  \label{fig:says-un-over-impl}
\end{figure*}

To see why, imagine that there are three principals of interest: Alice, Bob, and Cathy, and three labels: $\ell_{\textsf{P}}$, $\ell_{\textsf{S}}$, and $\ell_{\textsf{TS}}$, representing \pub, \secret, and \topsec, respectively.
(We use the shorter names to make our formal proofs easier to read.)
Anybody in the system can read public data (i.e., data labeled with $\ell_{\textsf{P}}$).
Alice and Cathy believe all three principals of interest can read secret data (i.e., data labeled with $\ell_{\textsf{S}}$), but Bob is unsure of the security clearances and will only send public data to other principals.
Alice and Cathy also have top secret clearance, but Bob does not, so he \emph{cannot} read data labeled at $\ell_{\textsf{TS}}$.
We can formalize these permission policies in the following context: \\
\parbox{\columnwidth}{
\begin{align*}
  \Gamma ={} & \belief{\forallexp{p}{\prin}{p \says*{\ell_{\textsf{S}}} \ell_{\textsf{P}} \flowsto \ell_{\textsf{S}}}}{\modalG{}}, \\
             & \belief{\forallexp{p}{\prin}{p \says*{\ell_{\textsf{TS}}} \ell_{\textsf{S}} \flowsto \ell_{\textsf{TS}}}}{\modalG{}}, \\
             & \belief{\canr(\Bob, \ell_{\textsf{S}})}{\modal{\Alice}{\ell_{\textsf{S}}}}, \\
             & \belief{\canr(\Alice, \ell_{\textsf{TS}})}{\modal{\Cathy}{\ell_{\textsf{TS}}}}, \\
             & \belief{\forallexp{p,q}{\prin}{p \says*{\ell_{\textsf{P}}} \canr(q, \ell_{\textsf{P}})}}{\modalG{}}, \\
             & \belief{\forallexp{p,q}{\prin}{\forallexp{\ell,\ell'}{\labl}{p \says*{\ell}{\canw(q,\ell')}}}}{\modalG{}}
\end{align*}}

Additionally, Bob serves as a redactor: given $\varphi$---which represents a document containing secret information---he can produce $\psi$---which represents a redacted version of the same document---performing a declassification in the process.
We represent Bob's ability by adding one belief:
$$\Gamma' = \Gamma, \belief{(\Bob \says*{\ell_{\textsf{S}}} \varphi) \mimpl* (\Bob \says*{\ell_{\textsf{P}}} \psi)}{\modalG{}}$$

\begin{figure*}
  \centering
  \begin{minipage}{\textwidth}
  $$\infer*[left=FwdL\supdag]{
    \infer*[Left=$\textsc{ImpL}'$\hspace{7.2em},leftskip=6.5em,rightskip=6.5em]{
      \infer*[Left=Ax]{ }{\Gamma, \belief{\varphi}{\modal{\Alice}{\ell_{\textsf{TS}}}} \proves \belief{\varphi}{\modal{\Alice}{\ell_{\textsf{TS}}}}}\\
      \infer*[Right=Ax]{ }{\Gamma, \belief{\varphi}{\modal{\Alice}{\ell_{\textsf{TS}}}}, \belief{\psi}{\modal{\Alice}{\ell_{\textsf{TS}}}} \proves \belief{\psi}{\modal{\Alice}{\ell_{\textsf{TS}}}}}
    }{\Gamma, \belief{(\varphi \mimpl* \psi)}{\modal{\Alice}{\ell_{\textsf{TS}}}}, \belief{\varphi}{\modal{\Alice}{\ell_{\textsf{TS}}}} \proves \belief{\psi}{\modal{\Alice}{\ell_{\textsf{TS}}}}}
  }{\Gamma, \belief{(\varphi \mimpl* \psi)}{\modal{\Alice}{\ell_{\textsf{TS}}}}, \belief{\varphi}{\modal{\Cathy}{\ell_{\textsf{TS}}}} \proves \belief{\psi}{\modal{\Alice}{\ell_{\textsf{TS}}}}}$$
  \end{minipage}
  \caption[Alice Redacts Cathy's Belief]{Alice using Cathy's $\varphi$ and a redaction function}
  \label{fig:using-alice-impl}
\end{figure*}

Imagine further that Alice decides she wants to redact secret information from a \topsec{} version of $\varphi$ that she receives from Cathy, but leave it \topsec{}.
If she can figure out how to get an implication representing redaction, she can simply receive $\varphi$ from Cathy and use the implication.
This is the proof in Figure~\ref{fig:using-alice-impl}.
For the sake of brevity and readability, we do not explicitly state side conditions that are proven straightforwardly from $\Gamma$.
The rules where these side conditions should appear are marked with~``\supdag.''

\begin{figure*}
  \begin{minipage}{\textwidth}
  $$
  \infer*[left=VarR\supdag,leftskip=1pt]{
    \infer*[Left=$\textsc{ImpR}'$,leftskip=1pt,rightskip=3.5pt]{
      \infer*[Left=VarR\supdag,leftskip=1pt,rightskip=3.5pt]{
        \infer*[Left=FwdL\supdag,leftskip=1pt,rightskip=3.5pt]{
          \infer*[Left=FwdR\supdag\hspace{10em},leftskip=9em,rightskip=9em]{
            \infer*[Left=$\textsc{ImpL}'$\hspace{2em},leftskip=1.8em,rightskip=2.3em]{
              \infer*[Left=SaysR]{
                \infer*[Left=Ax]{ }{\Gamma, \belief{\varphi}{\modal{\Bob}{\ell_{\textsf{S}}}} \proves \belief{\varphi}{\modal{\Bob}{\ell_{\textsf{S}}}}}
              }{\Gamma, \belief{\varphi}{\modal{\Bob}{\ell_{\textsf{S}}}} \proves \belief{\Bob \says*{\ell_{\textsf{S}}} \varphi}{\modalG{}}} \\
              \infer*[Right=SaysL]{
                \infer*[Right=Ax]{ }{\Gamma, \belief{\psi}{\modal{\Bob}{\ell_{\textsf{P}}}} \proves \belief{\psi}{\modal{\Bob}{\ell_{\textsf{P}}}}}
              }{\Gamma, \belief{\Bob \says*{\ell_{\textsf{P}}} \psi}{\modalG{}} \proves \belief{\psi}{\modal{\Bob}{\ell_{\textsf{P}}}}}
            }{\Gamma, \belief{(\Bob \says*{\ell_{\textsf{S}}} \varphi) \impl (\Bob \says*{\ell_{\textsf{P}}} \psi)}{\modalG{}}, \belief{\varphi}{\modal{\Bob}{\ell_{\textsf{S}}}} \proves \belief{\psi}{\modal{\Bob}{\ell_{\textsf{P}}}}}
          }{\Gamma', \belief{\varphi}{\modal{\Bob}{\ell_{\textsf{S}}}} \proves \belief{\psi}{\modal{\Alice}{\ell_{\textsf{P}}}}}
        }{\Gamma', \belief{\varphi}{\modal{\Alice}{\ell_{\textsf{S}}}} \proves \belief{\psi}{\modal{\Alice}{\ell_{\textsf{P}}}}}
      }{\Gamma', \belief{\varphi}{\modal{\Alice}{\ell_{\textsf{S}}}} \proves \belief{\psi}{\modal{\Alice}{\ell_{\textsf{S}}}}}
    }{\Gamma' \proves \belief{(\varphi \impl \psi)}{\modal{\Alice}{\ell_{\textsf{S}}}}}
  }{\Gamma' \proves \belief{(\varphi \impl \psi)}{\modal{\Alice}{\ell_{\textsf{TS}}}}}
  $$
\end{minipage}
\caption[Alice's Implication]{Proof corresponding to Alice sending $\varphi$ to Bob and receiving a $\psi$ back}
\label{fig:alice-implication}
\end{figure*}

While she knows how to \emph{use} an implication representing redaction, Alice does not know how to redact $\varphi$ except by giving it to Bob.
Using $\textsc{ImpL}'$ and $\textsc{ImpR}'$, she is able to package up the process ``give Bob a secret version of $\varphi$, get back a public version of $\psi$, and then use variance to get a secret version of $\psi$'' as a belief \mbox{$\belief{\varphi \mimpl* \psi}{\modal{\Alice}{\ell_{\textsf{S}}}}$}.
She can then use variance again to get a belief \mbox{$\belief{\varphi \mimpl* \psi}{\modal{\Alice}{\ell_{\textsf{TS}}}}$}.
This is the proof in Figure~\ref{fig:alice-implication}.
Again, we elide side conditions that are proven straightforwardly from $\Gamma$, and mark the rules where they should appear with ``\supdag.''

Cutting these two proofs together gives Alice what she wants: a \topsec{} version of $\psi$.
However, this cut is not possible to eliminate!
Examining this through a propositions-as-types lens tells us why: one of Alice or Cathy must send a \topsec{} version of $\varphi$ to Bob, which neither is willing to do.


%% file: signed-subformla.tex
Figure~\ref{fig:signed-subformula} contains the rules for the \emph{signed subformula relation}.

\begin{figure}
  \small
\begin{mathpar}
s \in \{+,-\} \and \overline{+} = - \and \overline{-} = + \\
\infer{ }{\varphi^s \leq \varphi^s} \and \infer{\varphi^s \leq \psi^{s'}\\ \psi^{s'} \leq \chi^{s''}}{\varphi^s \leq \chi^{s''}} \and
\infer{ }{\varphi^s \leq (\varphi \lor \psi)^s} \and \infer{ }{\psi^s \leq (\varphi \lor \psi)^s} \and \infer{ }{\varphi^s \leq (\varphi \land \psi)^s} \and \infer{ }{\psi^s \leq (\varphi \land \psi)^s} \and
\infer{ }{\varphi^{\overline{s}} \leq (\varphi \to \psi)^s} \and \infer{ }{\psi^s \leq (\varphi \to \psi)^s} \and
\infer{ }{(\subst{\varphi}{x}{t})^- \leq (\forallexp*{\varphi})^-} \and \infer{ }{\varphi^+ \leq (\forallexp*{\varphi})^+} \and
\infer{ }{\varphi^- \leq (\existsexp*{\varphi})^-} \and \infer{ }{(\subst{\varphi}{x}{t})^+ \leq (\existsexp*{\varphi})^+} \and
\infer{ }{\varphi^s \leq (p \says*{\ell} \varphi)^s} \and
\end{mathpar}

\caption{Signed Subformula Relation}
\label{fig:signed-subformula}
\end{figure}

Note that every subformula of a signed formula has a unique sign.
If a subformula appears by itself in a sequent during a proof, then which side of the turnstile it is on is determined by its sign.
This structure results in the following formal property.

\begin{thm}[Left Signed-Subformula Property]
  If $\Gamma \proves \belief{\varphi}{g_1}$ appears in a proof of $\Delta \proves \belief{\psi}{g_2}$,
  then for all $\belief{\chi_1}{g_3} \in \Gamma$, either (1) $\chi_1^- \leq \psi^+$ or (2) there is some $\belief{\chi_2}{g_4} \in \Delta$ such that $\chi_1^- \leq \chi_2^-$.
\end{thm}

This proof follows by induction on the FLAFOL proof rules.

Many logics also have a similar \emph{right} signed-subformula property.
FLAFOL does not enjoy that property since \mbox{$\Gamma \proves \belief{\varphi}{g_1}$} may be a side condition on a forward or a variance rule, and thus not related directly to~$\psi$.


%% file: simulation.tex
In (multi-)modal logics, we are interested in modeling \emph{perfect} reasoners.
That is, reasoners should reason correctly based on their assumed beliefs; if their assumed beliefs were true, then all of their derived beliefs would be as well.

In most logics (which do not have generalized principals, this is axiomatized as a rule in the system, written as follows:
$$\infer{\Gamma \proves \varphi}{p \says*{\ell} \Gamma \proves p \says*{\ell} \varphi}$$
Here, $p \says*{\ell} \Gamma$ refers to a copy of $\Gamma$ with $p \says*{\ell}$ in front of every formula in $\Gamma$.
In such logics, this is the main rule for manipulating says statements.
However, this requires removing all beliefs that are not those of $p$ at level $\ell$ in a context before using this rule to reason as $p$ at level $\ell$.

FLAFOL instead uses the $\says$ introduction rules in Section~\ref{sec:proof-system}, which allows us to retain the beliefs of other principals and of $p$ at other labels, making it easier to discuss communication.
This difference causes no harm.
FLAFOL reasoners are still be perfect reasoners, which we show by proving a theorem analogous to the above rule.
We refer to this as the \emph{simulation} theorem, since it says that $p$ is correctly simulating the world in its head.

Adopting the above rule directly fails for two reasons.
The first is that our belief syntax pushes $\says$ statements into generalized principals, so we must place the new principal-label pair at the beginning of the generalized principal instead of on the formula.
The second is that the semantics of implications in FLAFOL mean that \mbox{$p \says*{\ell} (\varphi \mimpl* \psi)$} has different semantics from \mbox{$(p \says*{\ell} \varphi) \mimpl* (p \says*{\ell} \psi)$}.
To address this concern, we define the $\odot$ operator:
$$\mbox{ $\modal{p}{\ell} \odot  \varphi \triangleq \begin{cases}
    (p \says*{\ell} (\modal{p}{\ell} \odot \psi)) \mimpl* (\modal{p}{\ell} \odot \chi) & \varphi = \psi \mimpl* \chi\\
                              (\modal{p}{\ell} \odot \psi) \land (\modal{p}{\ell} \odot \chi) & \varphi = \psi \land \chi\\
                              (\modal{p}{\ell} \odot \psi) \lor (\modal{p}{\ell} \odot \chi) & \varphi = \psi \lor \chi\\
                              \forallexp*{(\modal{p}{\ell} \odot \psi)} & \varphi = \forallexp*{\psi} \\
                              \existsexp*{(\modal{p}{\ell} \odot \psi)} & \varphi = \existsexp*{\psi} \\
                              \varphi & \text{otherwise}
\end{cases}$}$$
This essentially ``repairs'' implications to have the right $\says$ statements in front of the premise.

Because FLAFOL can move $\says$ statements into generalized principals, we need to lift the operator to beliefs.
In doing so, we must place $\modal{p}{\ell}$ at the \emph{beginning} of the generalized principal, leading to the following definition:
$$\modal{p}{\ell} \odot (\belief{\varphi}{\modalG{} \cdot g'}) \triangleq \belief{(\modal{p}{\ell} \odot \varphi)}{\modalG{} \cdot \modal{p}{\ell} \cdot g'},$$
From there we can lift the operator to contexts as well.
$$\mbox{$\modal{p}{\ell} \odot \Gamma \triangleq \begin{cases}
    \cdot & \Gamma = \cdot\\
    \left(\modal{p}{\ell} \odot \Gamma'\right), \modal{p}{\ell} \odot (\belief{\varphi}{g}) & \Gamma = \Gamma', \belief{\varphi}{g}
\end{cases}$}$$

With this definition in hand, we can now state the simulation theorem in full:
\begin{thm}[Simulation]
  \label{thm:simulation}
  The following rule is admissible:
  $$\infer{\Gamma \proves \belief{\varphi}{g}}{(\modal{p}{\ell} \odot \Gamma) \proves \modal{p}{\ell} \odot (\belief{\varphi}{g})}$$
\end{thm}


%% file: non-interference.tex
\section{Non-Interference}
\label{sec:non-interference}

Both authorization logics and information flow systems have important security properties called \emph{non-interference}~\cite{denning-lattice,GM82,garg2006non}.
On the face, these two notions of non-interference look very different, but their core intuitions are the same.
Both statements aim to prevent one belief or piece of data from interfering with another---\emph{even indirectly}---unless the security policies permit an influence.
Authorization logics traditionally define trust relationships between principals and non-interference requires that $p$'s beliefs affect the provability of $q$'s beliefs only when $q$ trusts $p$.
Information flow control systems generally specify policies as labels on program data and use the label flows-to relation to constrain how inputs can affect outputs.
For non-interference to hold, changing an input with label $\ell_1$ can only alter an output with label $\ell_2$ if $\ell_1 \flowsto \ell_2$.

FLAFOL views both trust between principals and flows between labels as ways to constrain communication of beliefs.
The forward rules model an authorization-logic-style sending of beliefs from one principal to another based on their trust relationships.
The label variance rules model a single principal transferring beliefs between labels based on the flow relationship between them.
By reasoning about generalized principals, which include both the principal and the label, we are able to capture both at the same time.
The result (Theorem~\ref{thm:NI}) mirrors the structure of existing authorization logic non-interference statements~\cite{garg2006non,abadi06}.
No similar theorem reasons about information flow or applies to policies combining discoverable trust and logical disjunction.
Theorem~\ref{thm:NI} does both.

\subsection{Trust in FLAFOL}

Building a notion of trust on generalized principals requires us to consider both the trust of the underlying (regular) principals and label flows.
The explicit label flow relation ($\flowsto$) cleanly captures restrictions on changing labels.
Trust between principals requires more care.
Alice may trust Bob with public data, but that does not mean she trusts him with secret data.
Similarly, Alice may believe that Bob can influence low integrity data without believing Bob is authorized to influence high integrity data.
This need to trust principals differently at different labels leads us to define our trust in terms of the two permission relations: $\canr(p, \ell)$ and $\canw(p, \ell)$.

We group label flows and principal trust together in a meta-level statement relating generalized principals.
As this relation is the fundamental notion of trust in FLAFOL, we follow existing authorization logic literature and call it \emph{speaks for}.

The speaks-for relation captures any way that one generalized principal's beliefs can be safely transferred to another.
This can happen through flow relationships ($g \cdot \modal{p}{\ell}$ speaks for $g \cdot \modal{p}{\ell'}$ if $\ell \flowsto \ell'$),
forwarding ($g \cdot \modal{p}{\ell}$ speaks for $g \cdot \modal{q}{\ell}$ if $p$ can forward beliefs at $\ell$ to $q$), and introspection ($g \cdot \modal{p}{\ell}$ speaks for $g \cdot \modal{p}{\ell} \cdot \modal{p}{\ell}$ and vice versa).
We formalize speaks-for with the rules in Figure~\ref{fig:spksfr-rules}.

\begin{figure}
  \centering
  \small
  \begin{mathpar}
    \infer*[left=ReflSF]{ }{\Gamma \proves g \spksfr g}
    \and
    \infer*[left=ExtSF]{\Gamma \proves g_1 \spksfr g_2}{\Gamma \proves g_1 \cdot \modal{p}{\ell} \spksfr g_2 \cdot \modal{p}{\ell}}
    \\
    \infer*[left=SelfLSF]{ }{\Gamma \proves g \cdot \modal{p}{\ell} \spksfr g \cdot \modal{p}{\ell} \cdot \modal{p}{\ell}}
    \and
    \infer*[left=SelfRSF]{ }{\Gamma \proves g \cdot \modal{p}{\ell} \cdot \modal{p}{\ell} \spksfr g \cdot \modal{p}{\ell}}
    \and
    \infer*[left=VarSF]{\Gamma \proves \belief{\ell \flowsto \ell'}{g \cdot \modal{p}{\ell'}}}{\Gamma \proves g \cdot \modal{p}{\ell} \spksfr g \cdot \modal{p}{\ell'}}
    \and
    \infer*[left=FwdSF]{
      \Gamma \proves \belief{\canr(q, \ell)}{g \cdot \modal{p}{\ell}} \\
      \Gamma \proves \belief{\canw(p, \ell)}{g \cdot \modal{q}{\ell}}
    }{\Gamma \proves g \cdot \modal{p}{\ell} \spksfr g \cdot \modal{q}{\ell}}
    \and
    \infer*[left=TransSF]{
      \Gamma \proves g_1 \spksfr g_2 \\
      \Gamma \proves g_2 \spksfr g_3
    }{\Gamma \proves g_1 \spksfr g_3}
  \end{mathpar}
  \caption{The rules defining speaks for.}
  \label{fig:spksfr-rules}
\end{figure}

To validate this notion of trust, we note that existing authorization logics often define speaks-for as an atomic relation
and create trust by requiring that, if $p$ speaks for $q$, then $p$'s beliefs can be transferred to $q$.
As our speaks-for relation exactly mirrors FLAFOL's rules for communication, it enjoys this same property.

\begin{thm}[Speaks-For Elimination]
  \label{SFElim}
  The following rule is admissible in FLAFOL:
  $$\infer*[left=ElimSF]{
    \Gamma \proves \belief{\varphi}{g_1} \\
    \Gamma \proves g_1 \spksfr g_2
  }{\Gamma \proves \belief{\varphi}{g_2}}$$
\end{thm}

With this notion of trust we can begin structuring a non-interference statement.
We might like to say that beliefs of $g_1$ can only influence beliefs of $g_2$ if $\Gamma \proves g_1 \spksfr g_2$,
or formally: if $\Gamma, (\belief{\varphi}{g_1}) \proves \belief{\psi}{g_2}$ is provable, then either $\Gamma \proves \belief{\psi}{g_2}$ is provable or $\Gamma \proves g_1 \spksfr g_2$.
Unfortunately, this statement is false for three critical reasons: $\says$ statements, implication, and the combination of discoverable trust and disjunctions.

\subsection{Says Statements and Non-Interference}

The first way to break the proposed non-interference statement above is simply by moving affirmations of a statement between the formula---using $\says$---and the generalized principal who believes it.
For example, we can trivially prove $\belief{p \says*{\ell} \varphi}{\modalG{}} \proves \belief{\varphi}{\modalG{} \cdot \modal{p}{\ell}}$, yet we cannot prove $\modalG{} \spksfr \modalG{} \cdot \modal{p}{\ell}$.

To address this case, we can view $\belief{p \says*{\ell} \varphi}{\modalG{}}$ as a statement that $\modalG{} \cdot \modal{p}{\ell}$ believes $\varphi$.
This insight suggests generally pushing all $\says$ modalities into the generalized principal.
We can do this for simple formulae, but the process breaks down with conjunction and disjunction.
In those cases, the different sides may have different $\says$ modalities, and either side could influence a belief through the different resulting generalized principals.
We alleviate this concern by considering a \emph{set} of generalized principals referenced in a given belief.
We build this set using an operator $\inmodals$:
$$\mbox{\small $\inmodals(\belief{\chi}{g}) \triangleq \begin{cases}
  \inmodals(\belief{\varphi}{g \cdot \modal{p}{\ell}}) & \chi = p \says*{\ell} \varphi \\
  \inmodals(\belief{\varphi}{g}) \cup \inmodals(\belief{\psi}{g}) & \chi = \varphi \land \psi \text{ or } \varphi \lor \psi \\
  \inmodals(\belief{\psi}{g}) & \chi = \varphi \mimpl* \psi \\
  \bigcup_{t : \sigma} \inmodals(\belief{\subst{\varphi}{x}{t}}{g}) & \chi = \forallexp*{\varphi} \text{ or } \existsexp*{\varphi} \\
  \{g\} & \text{otherwise}
\end{cases}$}$$
For implications, $\inmodals$ only considers the consequent, since only its consequent can affect the provability of a belief.
For quantified formulae, a proof may substitute any term of the correct sort for the bound variable, so we must as well.

Using this new operator, we can patch the hole $\says$ statements created in our previous non-interference statement, producing the following:
If $\Gamma, (\belief{\varphi}{g_1}) \proves \belief{\psi}{g_2}$, then either $\Gamma \proves \belief{\psi}{g_2}$, or there is some $g_1' \in \inmodals(\belief{\varphi}{g_1})$, $g_2' \in \inmodals(\belief{\psi}{g_2})$, and some $g_1''$ such that $\Gamma \proves g_1' \cdot g_1'' \spksfr g_2'$.

Here $g_1''$ represents the ability of a generalized principal to ship entire simulations to other generalized principals.
In particular, the forward and variance rules operate on an ``active'' prefix of the current generalized principal; $g_1''$ represents the ``inactive'' suffix.

The $\inmodals$ operator converts reasoning about beliefs from the object level (FLAFOL formulae) to the meta level (generalized principals).
FLAFOL's ability to freely move between the two forces us to push all such reasoning in the same direction to effectively compare the reasoner in two different beliefs.
Prior authorization logics do not contain a meta-level version of $\says$, meaning similar conversions do not even make sense.

\subsection{Implications}

While use of the $\inmodals$ function solves part of the problem with our original non-interference proposal, it does not address all of the problems.
Implications can implicitly create new trust relationships, allowing beliefs of one generalized principal to affect beliefs of another, even when no speaks-for relationship exists.
To understand how this can occur, we revisit our example of preventing SQL injection attacks from from Section~\ref{sec:prev-sql-inject}.

Recall from Section~\ref{sec:prev-sql-inject} that a web server might treat sanitized versions of low-integrity input as high integrity.
Further recall, it might represent this willingness with the following implication.
$$\sys \says*{\untrusted} \Input(x) \impl \sys \says*{\trusted} \Input(\san(x))$$

In an intuitively-sensible context where $\sys$ believes $\trusted \flowsto \untrusted$---high integrity flows to low integrity---but not vice versa, there is no way to prove $\modal{\sys}{\untrusted} \spksfr \modal{\sys}{\trusted}$.
The presence of this implication, however, allows some beliefs at $\modal{\sys}{\untrusted}$ to influence beliefs at $\modal{\sys}{\trusted}$.
This influence is actually an endorsement from $\untrusted$ to $\trusted$, and our speaks-for relation explicitly does not capture such effects.

Prior work manages this trust-creating effect of implications either by claiming security only when all implications are provable~\cite{abadi06} or by explicitly using assumed implications to represent trust~\cite{garg2006non}.
We hew closer to the latter model and make the implicit trust of implications explicit in our statement of non-interference.
We therefore cannot use the speaks-for relation, so we construct a new relation between generalized principals we call \emph{can influence}.

Intuitively, $g_1$ can influence $g_2$---which we denote \mbox{$\Gamma \proves g_1 \influences g_2$}---if either $g_1$ speaks for $g_2$ or there is an implication in $\Gamma$ that allows a belief of $g_1$ to affect the provability of a belief of $g_2$.
This relation, formally defined in Figure~\ref{fig:can-influence}, uses the $\inmodals$ operator discussed above to capture the generalized principals actually discussed by each subformula of the implication.
Because FLAFOL interprets the premise of an implication as a condition whose modality is independent of the entire belief, so too does the can-influence relation.
The relation is also transitive, allowing it to capture the fact that a proof may require many steps to go from a belief at $g_1$ to a belief at $g_2$.

\begin{figure}
  \centering
  \small
  \begin{mathpar}
    \infer*[left=SF-CI]{\Gamma \proves g_1 \spksfr g_2}{\Gamma \proves g_1 \influences g_2}
    \and
    \infer*[left=ExtCI]{\Gamma \proves g_1 \influences g_2}{\Gamma \proves g_1 \cdot g' \influences g_2 \cdot g'}
    \and
    \infer*[left=TransCI]{\Gamma \proves g_1 \influences g_2 \\ \Gamma \proves g_2 \influences g_3}{\Gamma \proves g_1 \influences g_3}
    \and
    \infer*[left=ImpCI]{
      \belief{\varphi \mimpl* \psi}{g} \in \Gamma \\
      g_1 \in \inmodals(\belief{\varphi}{\modalG{\ell}}) \\
      g_2 \in \inmodals(\belief{\psi}{g})
    }{\Gamma \proves g_1 \influences g_2}
  \end{mathpar}
  \caption{The rules defining the \emph{can influence} relation.}
  \label{fig:can-influence}
\end{figure}

Simply taking our attempted non-interference statement from above and replacing speaks-for with can-influence allows us to straightforwardly capture the effect of implications on trust within the system.

While this change may appear small, it results in a highly conservative estimate of possible influence.
Implications are precise statements that can allow usually-disallowed information flows under very particular circumstances.
Unfortunately, because our non-interference statement only considers the generalized principals involved, not the entire beliefs, it cannot represent the same level of precision.
A single precise implication added to a context can therefore relate whole classes of previously-unrelated generalized principals, eliminating the ability for non-interference to say anything about their relative security.
A similar lack of precision in information flow non-interference statements has resulted in long lines of research on how to precisely model or safely restrict declassification and endorsement~\cite{zm01b,sm04,ms04,lz05,ss05,msz06,cm08,am11,liopriv,nmifc}.

\subsection{Discovering Trust with Disjunctions}
\label{sec:disc-trust-disj}

The $\inmodals$ operator and can-influence relation address difficulties from both $\says$ formulae and implications, but our statement of non-interference still does not account for the combination of disjunctions and the ability to discover trust relationships.
To understand the effect of these two features in combination, recall the reinsurance example from Section~\ref{sec:prov-hosp-bill}.
Bob can derive \mbox{$\canw(I_1, \ell_H)$} if he already believes both \mbox{$\canw(I_1, \ell_H) \mathrel{\lor} \canw(I_2, \ell_H)$} and \mbox{$I_2 \says*{\ell_H} \canw(I_1, \ell_H)$}.
We clearly cannot remove either of Bob's beliefs and still prove the result.
Our desired theorem statement would thus require that $\modal{\Bob}{\ell_H} \cdot \modal{I_2}{\ell_H}$ can influence $\modal{\Bob}{\ell_H}$, which there is no way to prove.
The reason the sequent is still provable, as we noted in Section~\ref{sec:prov-hosp-bill}, is that Bob can \emph{discover} trust in $I_2$ when he branches on an Or statement, which then allows $I_2$ to influence Bob.
In this branch, we can prove $\modal{\Bob}{\ell_H} \cdot \modal{I_2}{\ell_H} \spksfr \modal{\Bob}{\ell_H} \cdot \modal{\Bob}{\ell_H}$, which then speaks for $\modal{\Bob}{\ell_H}$.

To handle such assumptions, we cannot simply consider the context in which we are proving a sequent;
we must consider any context that can appear in the proof of that sequent.
We developed the notion of compatible supercontexts in Section~\ref{sec:compat-supercontext} for exactly this purpose.
Indeed, if we replace $\Gamma$ with an appropriate CSC when checking the potential influence of generalized principals, we remove the last barrier to a true non-interference theorem.

\subsection{Formal Non-Interference}

The techniques above allow us to modify our attempted non-interference statement into a theorem that holds.
\begin{thm}[Non-Interference]
  \label{thm:NI}
  For all contexts $\Gamma$ and beliefs $\belief{\varphi}{g_1}$ and $\belief{\psi}{g_2}$, if
  $$\Gamma, \belief{\varphi}{g_1} \proves \belief{\psi}{g_2},$$
  then either (1) $\Gamma \proves \belief{\psi}{g_2}$, or (2) there is some \mbox{$\Delta \ll \Gamma, \belief{\varphi}{g_1} \proves \belief{\psi}{g_2}$}, $g_1' \in \inmodals(\belief{\varphi}{g_1})$, $g_2' \in \inmodals(\belief{\psi}{g_2})$, and $g_1''$
  such that $\Delta \proves g_1' \cdot g_1'' \influences g_2'$.
\end{thm}

The proof of this theorem follows by induction on the proof of \mbox{$\Gamma, \belief{\varphi}{g_1} \proves \belief{\psi}{g_2}$}.
For each proof rule, we argue that either $\belief{\varphi}{g_1}$ is unnecessary for all premises or we can extend an influence from one or more subproofs to an influence from $\belief{\varphi}{g_1}$ to $\belief{\psi}{g_2}$.

This theorem limits when a belief $\belief{\varphi}{g_1}$ can be necessary to prove $\belief{\psi}{g_2}$ in context $\Gamma$, much like other authorization logic non-interference statements~\cite{garg2006non,abadi06}.
As we mentioned above, however, it is the first such non-interference statement for any authorization logic supporting all first-order connectives and discoverable trust.
Moreover, it describes how FLAFOL mitigates both:
\begin{itemize}
  \item communication between principals, through $\canr$ and $\canw$ statements, and
  \item movement of information between security levels represented by information flow labels, via flows-to statements.
\end{itemize}

The $\influences$ relation seems to make our non-interference statement much less precise than we would like.
After all, implications precisely specify what beliefs can be declassified or endorsed, whereas $\influences$ conservatively assumes any beliefs can move between the relevant generalized principals.
This lack of precision serves a purpose.
It allows us to reason about any implications, including those that arbitrarily change principals and labels, something which other no authorization logics have done before.
It is therefore worth noting that, when all of the implications in the context are provable, the theorem holds \emph{even if you replace $\influences$ with $\spksfr$ everywhere.}
The same proof works, with some simple repair in the \textsc{ImpL} case.

Another complaint of imprecision applies to compatible supercontexts.
Specifically, if any principal assumes $\varphi \lor \neg \varphi$ for any formula $\varphi$, then there is a CSC in which that principal has assumed both, even though these are arrived at through mutually-exclusive choices.
Since CSCs have been added in order to allow disjunctions and discoverable trust to co-exist, it is good to know that if we disallow either, CSCs are not required for non-interference.
That is, if there are no disjunctions in the context, then we can always instantiate the $\Delta$ in Theorem~\ref{thm:NI} with $\Gamma, \belief{\varphi}{g_1}$.
Similarly, if every permission that is provable in any CSC of $\Gamma, \belief{\varphi}{g_1} \proves \belief{\psi}{g_2}$ is provable under $\Gamma, \belief{\varphi}{g_1}$, then we can again always instantiate $\Delta$ with $\Gamma, \belief{\varphi}{g_1}$.

Together, these points demonstrate that there are only two types of poorly-behaved formulae that force the imprecision in Theorem~\ref{thm:NI}.
This further shows that our non-interference result is no less precise than those of other authorization logics in the absence of such formulae.
We add imprecision only when needed to allow our statement to apply to more proofs. 

To see how Theorem~\ref{thm:NI} corresponds to traditional non-interference results for information flow, consider a setting where every principal agrees on the same label ordering, and where there are no implications corresponding to declassifications or endorsements.
Then any two contexts~$\Gamma$ and~$\Gamma'$ which disagree only on beliefs labeled above some $\ell$ can prove exactly the same things at label $\ell$---$\Gamma \proves \belief{\varphi}{g \cdot \modal{p}{\ell}}$ if and only if $\Gamma' \proves \belief{\varphi}{g \cdot \modal{p}{\ell}}$---since Theorem~\ref{thm:NI} allows us to delete all of the beliefs on which they disagree.
If we view contexts as inputs, as in a propositions-as-types interpretation, then this says that changing high inputs cannot change low results.


%% file: future.tex
\section{Future Work}
\label{sec:future-work}

FLAFOL is already very powerful, but it suggests numerous avenues for future work.

First, FLAFOL only disallows \emph{direct} flows of information in proofs, but checking proofs can cause communication and potentially leak information.
Importantly, eliminating cuts in proofs can \emph{increase} the information leaked during proof-checking because eliminating cuts can reduce the uncertainty about which discoveries can be made during a proof.
This is disturbing, since we would like to be able to perform sound security analyses on proofs with cut;
system designers should not need to understand the very complicated cut-elimination proof.
The \emph{program counter} mechanism used by information flow control systems like Fabric~\cite{jfabric} and FLAM~\cite{flam} seems to prevent similar leaks.
Incorporating program counter labels to limit communication in FLAFOL proofs could eliminate these leaks in FLAFOL as well.

This improvement also widens the range of programs that can safely use FLAFOL.
Justifications for authorization need to be found as well as checked.
From the point of view of an authorization logic, this corresponds to proof search.
Searching for an authorization proof in a distributed system, however, may require communication between principals, potentially leaking why they are searching for this proof in the first place.
One  avenue forward embeds FLAFOL in a language with information-flow types, and runs proof search in that language.
This would guarantee that the proof search does not leak data assuming FLAFOL proofs do not leak data when checked.

We have developed new techniques to reason about authorization-logic proofs in order to prove non-interference for FLAFOL.
These reasoning principles could be expanded and used in other logics.
For instance, using the tools developed in Section~\ref{sec:non-interference}, we should be able to give non-interference proofs for logics like NAL~\cite{NAL} and FOCAL~\cite{HirschClarkson} which reason about implication and disjunction.
We should also be able to add disjunction and implication to logics like DCC~\cite{abadi06,ccd99} while still providing a non-interference theorem.

Another avenue of further work would understand better how $\says$ statements can interact with other logical connectives.
For instance, one might want to model a principal who cannot observe whether they are holding evidence of $\varphi$ or of $\psi$.
For instance, we might want to model a principal $p$ who receives an encrypted message containing a bit $b$.
Then $p$ knows that either $b = 0$ or $b = 1$, but $p$ has no way to determine which.
Thus, while $p \says*{\ell} (b = 0 \lor b = 1)$, we should not be able to show that $(p \says*{\ell} b = 0) \lor (p \says*{\ell} b = 1)$.
A NuPRL-like ``squash'' operator, which prevents evidence from being used~\cite{Caldwell}, could model this,
but further research is needed for FLAFOL to reason about the security of such protocols.

A similar avenue for future work involves exploring ways to allow $\says$ to distribute over implications while remaining coherent.
One potential approach would be to confine most reasoning to a single generalized principal, but this would restrict implications so that the principal who believes them cannot communicate in their proof.
The consequences of such a restriction on modeling real-world systems are unclear.

Finally, it would be nice to reason about the \emph{temporal} components of authorization; this is one place where work on information flow far outstrips that on authorization logic~\cite{zm01b,sm04,ms04,lz05,ss05,msz06,cm08,am11,liopriv,nmifc}.
Trust relationships may change over time, allowing or disallowing communication pathways.
Understanding how this changes which authorizations should be provable, and how this affects information-flow policies, is a rich area for exploration.


%% file: related.tex
\section{Related Work}
\label{sec:related}

Prior work in information flow and authorization logics has explored the connection between the two.
The Decentralized Label Model~\cite{ml-sp98,ml-tosem} includes a notion of ownership in information flow policies specifying who may authorize exceptions to the policy.
The Flow-Limited Authorization Model~(FLAM)~\cite{flam} was the first logic to directly consider the effects of data confidentiality and integrity on trust relationships between principals.
Prior work on Rx~\cite{shtz06} and RTI~\cite{RTI} enforced language-based information flow policies via \emph{roles} whose membership were protected with confidentiality and integrity labels.
By contrast, FLAFOL is a formal authorization logic containing every first-order connective.

\paragraph{Decentralized Label Model}
The Decentralized Label Model~(DLM)~\cite{ml-sp98,ml-tosem} is a model for expressing information flow labels in a decentralized system.
Its labels contain two components, confidentiality and integrity, which are each specified as a set of principals who may read or write the data, respectively.
FLAFOL separates principals and labels by making them independent sorts that are related by the $\canr$ and $\canw$ relations.
This allows system designers much more freedom in determining the semantics of principals and labels.
For instance, DLM labels cannot represent availability.

DLM labels also include a notion of ownership, but it only specifies who may authorize exceptions to the policy.
FLAFOL has no built-in ownership notion, nor does it allow specific exceptions to policies.

Moreover, DLM assumes that labels form a global static lattice.
As we have discussed in detail, FLAFOL does not make this assumption.
In particular, FLAFOL labels need not be static, since they can be the result of functions.
Second, they need not be a lattice, but merely a partial order.
Finally, the order of FLAFOL labels need not be global, since different (generalized) principals may have very different ideas of the order.
This generality allows FLAFOL to be extremely expressive, as we saw in Section~\ref{sec:system-model}.

\paragraph{Flow-Limited Authorization Model}
The Flow-Limited Authorization Model~(FLAM)~\cite{flam} was the first information-flow label model to directly consider the interaction between information flow and authorization.
FLAM does not, however, provide a full authorization logic.
It lays out important rules for reasoning about communication in systems with discoverable trust relationships where principals may disagree on those relationships.
It also restricts participation in a proof using a program counter label to help full systems remain secure in contexts where merely checking a proof may leak data.
FLAM, however, provides no means to directly express authorization policies other than one principal trusting another.
It has no first-order connectives or quantifiers and no way for one principal to reason about another's beliefs.

FLAM also takes the principal-label connection a step beyond the DLM and represents principals directly as a combination of confidentiality and integrity labels.
This view restricts FLAM from reasoning about labels with policies other than confidentiality and integrity, since they might necessitate subtle changes to FLAM's reasoning rules.
FLAFOL's $\canr$ and $\canw$ relations abstract out how different label components may interact, allowing each system to specify appropriate restrictions given the meaning of its labels.

Unifying principals and labels also undermines FLAM's effectiveness as an authorization logic.
It is often convenient to construct complex policies from simpler ones, such as a policy protecting Alice's confidentiality and Bob's integrity.
FLAM regards such a compound policy as a principal, breaking the connection between formal principals and system entities.
While FLAFOL can certainly represent these policies, doing so does not force a reasoner to break this connection.

FLAM additionally does not provide a non-interference guarantee, instead offering a guarantee called \emph{robust authorization}.
In FLAM, each fact has a label representing its confidentiality and integrity and is stored on a node, which is itself represented by a label.
If a node $c$ believes a derived fact at label $\ell$, robust authorization says:
\begin{itemize}
  \item The label of every fact used in the derivation flows to $\ell$,
  \item Every node in the derivation may control whether the derivation took place,
  \item $c$ is allowed to learn every fact used in the derivation, and
  \item For each node~$n$ involved in the derivation, $c$ will listen to~$n$ at~$\ell$ and~$n$ will talk to~$c$ at~$\ell$.
\end{itemize}

FLAFOL's non-interference theorem gives similar guarantees.
In particular, our non-interference theorem shows that the label of every belief used in a derivation (without implications) flows to the label of the derived belief.
Moreover, for each belief \mbox{$\belief{\varphi}{g}$} used in a derivation without implications, the generalized principal who believes the conclusion must (transitively) trust~$g$.

However, FLAFOL does not have any notion of who may control whether a derivation takes place.
We are able to achieve FLAFOL's security guarantee without the restrictions imposed by FLAM's program counter label.

\paragraph{DCC and FLAC}
The Dependency Core Calculus~(DCC)~\cite{ccd99} is a small functional core calculus designed to capture dependencies within programs, including information flows.
It uses a monadic structure to represent labels at the level of types and enforce standard information flow typing constraints.
Abadi also reinterpreted DCC's type system as an authorization logic~\cite{abadi06}, but used the modalities created by the monadic structure to represent principals' beliefs.
This technique allows DCC to reason about either information flow or authorization, but not both at the same time.
DCC does provide a non-interference property, but it employs a static external lattice to express trust.

The Flow-Limited Authorization Calculus~(FLAC)~\cite{flac} builds a computational model for FLAM by extending Polymorphic DCC~\cite{abadi06} with discoverable trust relationships.
It uses DCC's information-flow interpretation and FLAM's discoverable trust rules to bound information flows and how they can affect trust assumptions.

Because FLAC incorporates DCC's computational model, we can view its type system as a propositional logic that reasons about discoverable trust.
Since the logic is based on System F, it contains some elements of second-order logic by supporting universal quantification over types, but lacks any existential quantification.
Critically, FLAC programs execute only on a single machine with no notion of communication.
This means that, unlike both FLAM and FLAFOL, it does not allow reasoning about the interaction between different system components with different trust assumptions, and thus does not form a full authorization logic.
It can only reason about how data may influence trust assumptions and resulting decisions within a single component.
DFLATE~\cite{dflate} extends FLAC with channels that support a limited form of communication.

FLAC provides strong information security guarantees for computations defined in the language.
The local nature of these computations, however, means these assurances apply only to local reasoning on a single host.
FLAFOL, by contrast, provides strong security guarantees in a fully distributed context when reasoning about differing beliefs within a system.
It does not yet have an associated programming model, but developing one would be interesting future work.

\paragraph{Other Authorization Logics}
Becker~\cite{becker2012information} explores preventing probing attacks, authorization queries which leak secret information, in Datalog-based authorization logics like DKAL~\cite{DKAL} and SecPAL~\cite{SecPAL}.
In SecPAL\textsuperscript{+}~\cite{secpal-plus}, Becker proposes a new \emph{can listen to} operator, similar to FLAFOL's $\canr$ permission, that expresses who is permitted to learn specific statements.
However, \emph{can listen to} expresses permissions on specific statements, not labels as $\canr$ does.
Moreover, FLAFOL tracks dependencies between statements using these labels, so the security consequences of adding a new permission are more explicit.

Garg and Pfenning~\cite{garg2006non} present an authorization logic and a non-interference result that ensures untrusted principals cannot influence the truth of statements made by other principals.
Garg and Pfenning, however, support a more limited set of logical connectives than FLAFOL, use only implications to encode trust, and do not reason directly about information flow.

Finally, \textsc{Aura}\cite{aura,aura-if} embeds DCC into a language with dependent types to explore how authorization logic interacts with programs.
They inherit their non-interference result directly from DCC, but they express first-order properties by combining other programming language constructs with DCC.
This makes it unclear what guarantees the theorem provides.
Jia and Zdancewic encode information-flow labels into \textsc{Aura} as principals and develop a non-interference theorem in the style of information-flow systems~\cite{aura-if}.
This setup unfortunately makes it impossible for principals to disagree about the meaning of labels, since the labels themselves define their properties.


%% file: conclusion.tex
\section{Conclusion}
\label{sec:conclusion}

We have introduced FLAFOL, a first-order logic which combines notions of trust from both authorization and information flow.
It provides a concrete model of communication that respects this combination
and gives principals the ability to reason about each other's differing opinions, including differing opinions about trust.
FLAFOL has a powerful non-interference theorem that navigates this complexity, a top-tier result for authorization logics.
It is, moreover, the most complete first-order logic with such a guarantee.


%% file: acknowledgements.tex
\section*{Acknowledgments}
\label{sec:acknowledgments}

We would first like to thank our anonymous reviewers for their insightful comments and helpful suggestions.
Andrew Myers and Jed Liu provided early feedback on the design of FLAFOL.
Discussion with Deepak Garg gave us insight into how to introduce FLAFOL, while Phokion G. Kolatis pointed us to some related work.
We would finally like to thank Co\c{s}ku Acay, Arthur Azevedo de~Amorim, Eric Campbell, Dietrich Geisler, Elisavet Kozyri, Tom Magrino, Matthew Milano, Andrew Morgan, and Drew Zagieboylo for their valuable help editing this paper.

Funding for this work was provided by NSF grant \#1704788, NSF CAREER grant \#1750060, and a National Defense Science and Engineering Graduate Fellowship.
Any opinions, findings, conclusions, or recommendations expressed here are those of the authors and may not reflect those of these sponsors.


%% file: examples_flafol.tex
\section{Formalizing the Examples}
\label{sec:formalizing-examples}

In Section~\ref{sec:flafol-example}, we discussed several examples of authorization policies that interact with information-flow policies in non-trivial ways.
While we used these to introduce FLAFOL's features and syntax, in Sections~\ref{sec:looking-at-pictures} and~\ref{sec:prov-hosp-bill} we elided some technical details in order to simplify the presentation.
In this section, we formalize these examples, making it clear how FLAFOL can represent each of the policies described in Section~\ref{sec:flafol-example}.

\subsection{Viewing Pictures on Social Media}
\label{sec:view-pict-soci}

Recall the example in Section~\ref{sec:looking-at-pictures}: Bob has uploaded a picture to a social-media account along with a policy that only those on a friend list that he maintains on that account may view the photo.
Moreover, he has a policy that only his friends may know who is on his friend list.

We mentioned that we can represent Bob's friend list as a collection of beliefs of the form $\Bob \says*{\frnd} \isFriend(p)$.
We can now discuss these beliefs in more detail.
Bob's friend list contains a finite number of principals; let $L$ be the set of principals on the list.
We can then represent Bob's friend list with the following FLAFOL assumptions:
$$\def\arraystretch{1.2}
\Gamma_L =
  \{\Bob \says*{\frnd} \isFriend(p) \mid p \in L\}
  \cup \left\{
    \begin{array}{@{~}l@{~}}\forallexp{q}{\prin}{} \\
      \quad\Bob \says*{\frnd} \isFriend(q) \\
      \quad\impl \left(\bigvee_{p \in L} \Bob \says*{\frnd} q = p\right)
    \end{array}\right\}$$
(For brevity, we omit ``$\atl \modalG{}$'' on all beliefs at~$\modalG{}$.)

This set of beliefs, along with some beliefs about equality (such as reflexivity, decidability of equality on principals, and Bob's belief that if two principals are equal, then one is a friend if the other is) are enough to determine Bob's friend list.
In particular, one can show that
$$\Gamma_L, \Gamma \proves \forallexp{p}{\prin}{\left(\begin{array}{@{}l@{}}
                                     \Bob \says*{\frnd} \isFriend(p)\\
                                     \lor~\Bob \says*{\frnd} \lnot \isFriend(p)
\end{array}\right)}.$$
This proof is conceptually simple, but quite tedious, so we elide it here.
This is a decision procedure since any FLAFOL proof of $\Gamma \proves \belief{\varphi \lor \psi}{g}$ can be transformed into a proof of either $\Gamma \proves \belief{\varphi}{g}$ or $\Gamma \proves \belief{\psi}{g}$.
\footnote{This has been proven in Coq as part of the cut-elimination proof.}

Recall that we created the label $\frnd$ in order to represent Bob's policy ``I will only share this with my friends.''
However, we never showed how to connect this with Bob's friend list, expressed as the relation $\isFriend(\prin)$.
In order to make this connection, we use a bi-implication
$$\forallexp{p}{\prin}{\left(\begin{array}{@{}l@{}}\Bob \says*{\frnd} \isFriend(p)\\ \leftrightarrow \Bob \says*{\frnd} \canr(p, \frnd)\end{array}\right)}$$

It might seem strange to have an implication from $\canr(p, \frnd)$ to $\isFriend(p)$.
After all, this seems to suggest that, if a principal can read Bob's $\frnd$ label, then Bob is going to be willing to consider them a friend, even if they were not on his friend list.
However, note that in $\Gamma_L$ we have a belief that says that if a principal is a friend, then they are one of the principals on Bob's friend list.
Thus, the troublesome implication above actually suggests that only those principals in $L$ can read the label $\frnd$.

While we can decide whether Bob believes that somebody is on his friend list---and therefore whether they can read things labeled $\frnd$---when Alice tries to look at Bob's picture the system needs to be able to tell Alice whether she is allowed to do so or not.
We informally argued earlier that this was impossible.
To see why, imagine that there is a label $\ell$ that we know Alice can read.
Since we don't know if Alice can read things of label $\frnd$, we assume that $\Gamma_L, \Gamma \nproves \belief{\frnd \flowsto \ell}{\modalG{} \cdot \modal{\Bob}{\ell}}$.
But then it is not the case that $\modalG{} \cdot \modal{\Bob}{\frnd} \spksfr \modalG{} \cdot \modal{\Bob}{\ell}$,
so the same holds for $\influences$ in the absence of irrelevant implications.
Now if we have any proof of
$$\Gamma_L, \Gamma \proves \forallexp{p}{\prin}{\left(\begin{array}{@{}l@{}}\Bob \says*{\ell} \isFriend(p)\\ \lor~\Bob \says*{\ell} (\lnot \isFriend(p))\end{array}\right)}$$
by Theorem~\ref{thm:NI}, we could remove Bob's friend list and get a proof under only $\Gamma$, which is clearly impossible.

As we discussed in Section~\ref{sec:looking-at-pictures}, this suggests that Bob should choose a more-permissive policy for his friend list.
One possibility is for Bob to label his friend list publicly, but this is not a very satisfying solution.
Another possibility is for Bob to allow any principal~$p$ to know whether~$p$ is on the list, but to not allow any principal~$p$ that is not on the list to know the status of any other principal.
However, we did not discuss how to represent this policy in FLAFOL.

One simple way to represent this policy is using an implication.
That is, we can assume the following:
\begin{align*}
  \forall {p} \mkern-3mu:\mkern2mu & \prin. \\
  & \left(\begin{array}{@{}l@{}}
        \Bob \says*{\frnd}\isFriend(p)\\
        \quad\to p \says*{\frnd}(\Bob \says*{\frnd}\isFriend(p))
      \end{array}\right) \\[0.2\baselineskip]
  & \land \left(\begin{array}{@{}l@{}}
        \Bob \says*{\frnd}\lnot\isFriend(p)\\
        \quad\to p \says*{\frnd}(\Bob \says*{\frnd}\lnot\isFriend(p))
      \end{array}\right)
\end{align*}
This allows $p$ to know whether $p$ is on Bob's friend list.
However, it is rather unsatisfying, because it destroys all of the security guarantees of Theorem~\ref{thm:NI}.
After all, we can now show that \mbox{$\modalG{} \cdot \modal{\Bob}{\frnd} \influences \modalG{} \cdot \modal{p}{\frnd}$} for any $p$.

We can create a more-subtle version of this policy which still enjoys the guarantees of Theorem~\ref{thm:NI}.
To do this, Bob labels his belief about whether or not a principal $p$ is on his friend list at a label $f(p)$ that $p$ may read and Bob's friends may read, but no one else.
We thus create a function symbol $f : \prin \to \labl$ and use it to define Bob's friend list.
Now we can re-define $\Gamma_L$ as follows:
$$\Gamma_L =
  \{\Bob \says*{f(p)} \isFriend(p) \mid p \in L\}
  \cup \left\{
    \begin{array}{@{~}l@{~}}\forallexp{q}{\prin}{} \\
      \quad\Bob \says*{f(q)} \isFriend(q) \\
      \quad\impl \left(\bigvee_{p \in L} \Bob \says*{f(q)} q = p\right)
    \end{array}\right\}$$
If $\Gamma$ contains axioms about equality, then
$$\Gamma_L, \Gamma \proves \forallexp{p}{\prin}{\left(\begin{array}{@{}l@{}}
                                     \Bob \says*{f(p)} \isFriend(p)\\
                                     \lor~\Bob \says*{f(p)} \lnot \isFriend(p)
\end{array}\right)}.$$

We now need to formalize the statement that $p$ and Bob's friends may read $f(p)$, but nobody else.
We do this with the following assumptions $\Gamma_f$.
$$\def\arraystretch{1.2}
\Gamma_f =
  \left\{\begin{array}{@{}l@{}}
    \forallexp{p}{\prin}{\Bob \says*{f(p)} \canr(p, f(p))}, \\
    \forallexp{p}{\prin}{\Bob \says*{\frnd} f(p) \flowsto \frnd},\\
     \forallexp{p}{\prin}{\left(\begin{array}{@{}l@{}}\Bob \says*{f(p)} \isFriend(p)\\ \leftrightarrow \Bob \says*{f(p)} \canr(p, \frnd)\end{array}\right)}
  \end{array}\right\}$$
These rules allow Bob to forward to $p$ the results of the above decision procedure on $\Bob \says*{f(p)} \isFriend(p)$ if $p$ will listen---which requires $p \says*{f(p)} \canw(\Bob, f(p))$.
Similarly, if $p \in L$, then Bob will forward whether or not $q$ is Bob's friend for any principal $q$ at label $\frnd$.

\subsection{Hospital Bills Calculation and Reinsurance}
\label{sec:hosp-bills-calc}

Recall the example from Section~\ref{sec:prov-hosp-bill}: Alice has two possible insurers, $I_1$ and $I_2$.
Bob is trying to figure out which will be allowed to influence Alice's hospital bill, labeled $\ell_H$.
He represents the fact that either $I_1$ or $I_2$ will be able to influence the bill as $\Bob \says*{\ell_H} (\canw(I_1, \ell_H) \lor \canw(I_2, \ell_H))$.
He knows that $I_2$ reinsures with $I_1$, which we represent as $\Bob \says*{\ell_H} (I_2 \says*{\ell_H} \canw(I_1, \ell_H))$.

In Section~\ref{sec:prov-hosp-bill}, we did not discuss the confidentiality requirements of this situation.
In this case, since both insurers know that Bob is a hospital administrator, they are willing to talk to Bob about $\ell_H$.
Bob moreover knows this.
Therefore, we can use \mbox{$\Bob \says*{\ell_H} (I_2 \says*{\ell_H} \canr(\Bob, \ell_H)),$} which we will need in the proof.

\begin{figure*}[tb]
  \begin{minipage}{\textwidth}
    \begin{mathpar}
      \infer*[left=OrL]{ 
        \infer*[Left=Ax]{ }{\Gamma', \varphi \proves \varphi}\\
        \infer*[Right=$\textsc{FwdL}\supdag$]{
          \infer*[Right=SelfL]{
            \infer*[Right=Ax]{ }{
              \Gamma', \belief{\canw(I_2, \ell_H)}{\modalG{} \cdot \modal{\Bob}{\ell_H}}, \belief{\canw(I_1, \ell_H)}{\modalG{} \cdot \modal{\Bob}{\ell_H}} \proves \varphi
            }}
          {
            \Gamma', \belief{\canw(I_2, \ell_H)}{\modalG{} \cdot \modal{\Bob}{\ell_H}}, \belief{\canw(I_1, \ell_H)}{\modalG{} \cdot \modal{\Bob}{\ell_H} \cdot \modal{\Bob}{\ell_H}} \proves \varphi
          }}
        {
          \Gamma', \belief{\canw(I_2, \ell_H)}{\modalG{} \cdot \modal{\Bob}{\ell_H}} \proves \varphi
        }}
      {
        \Gamma \proves \varphi
      }
    \end{mathpar}
  \end{minipage}
  \caption{Bob's proof that $I_1$ can influence the bill}
  \label{fig:I_1-can-infl}
\end{figure*}

We can then formalize this example as a proof of the sequent $\Gamma \vdash \belief{\canw(I_1, \ell_H)}{\modalG{} \cdot \modal{\Bob}{\ell_H}}$ where
$$\def\arraystretch{1.2}
\Gamma = \left\{\begin{array}{@{~}l@{~}}
              \belief{\left(\begin{array}{@{}l@{}}\canw(I_1, \ell_H)\\ \lor~\canw(I_2, \ell_H)\end{array}\right)}{\modalG{} \cdot \modal{\Bob}{\ell_H}},\\
               \belief{\canw(I_1, \ell_H)}{\modalG{} \cdot \modal{\Bob}{\ell_H} \cdot \modal{I_2}{\ell_H}},\\
               \belief{\canr(\Bob, \ell_H)}{\modalG{} \cdot \modal{\Bob}{\ell_H} \cdot \modal{I_2}{\ell_H}}
             \end{array}\right\}$$
Note that we have moved from $\says$ statements to generalized principals here.
This is conceptually the same, and simply requires less work to move all of the says statements to the generalized principal in the proof.
The formal proof is available in Figure~\ref{fig:I_1-can-infl}, where we use a few pieces of shorthand for brevity and readability.
First we refer to the belief \mbox{$\belief{\canw(I_1, \ell_H)}{\modalG{} \cdot \modal{\Bob}{\ell_H}}$} as $\varphi$.
Second, we let
$$\Gamma' = \left\{\begin{array}{l}
               \belief{\canw(I_1, \ell_H)}{\modalG{} \cdot \modal{\Bob}{\ell_H} \cdot \modal{I_2}{\ell_H}},\\
               \belief{\canr(\Bob, \ell_H)}{\modalG{} \cdot \modal{\Bob}{\ell_H} \cdot \modal{I_2}{\ell_H}}
             \end{array}\right\}$$
Finally, we do not explicitly state side conditions which are proven straightforwardly from $\Gamma$.
The rules where these side conditions should appear are marked with ``\supdag.''


%% file: permissionmodels.tex
\section{Examples of Permission Models}
\label{sec:exampl-perm-models}

In Section~\ref{sec:proof-system} we saw how FLAFOL can be used to reason about a capabilities-based system.
However, FLAFOL's flexibility allows it to model many other kinds of systems.
In this appendix, we explore modeling two other systems in FLAFOL: a simple system with no additional assumptions, and a system similar to military classification levels or FLAM's system.

There is no particular reason for there to be some external model of permissions.
The ``default'' permission model simply gives meaning to $\canr$ and $\canw$ through their behavior.
That is, the only properties FLAFOL assumes about $\canr$ and $\canw$ are variance constraints, while all other properties of $\canr$ and $\canw$ come from formulae in the context of a proof.
This is appropriate in many cases.
For instance, in the example of viewing photos on social media, $\canr$ and $\canw$ have their behavior tuned by Bob's selections on his account settings page.
It is appropriate for the behaviors based on the selections to be axiomatized directly, rather than forced into some other model.
Note that since we only care about confidentiality in that example, $\canw$ can have a trivial implementation:
$$p \says*{\ell'}\canw(q, \ell) \leftrightarrow \True.$$

FLAFOL can encode a more-concrete possible permission model by assigning every principal a label representing ``which data this person is allowed to read or write.''
This model appears in the real world in the U.S. military, where every person has a clearance label, and they are allowed to read documents labeled at or below their clearance.
A more subtle version of this model separates reading and writing into confidentiality and integrity labels and allows every principal to have their own idea of each person's label.
This is similar to FLAM's model, though our version is typed and does not force principals and labels to be the same.

We can formalize this by giving projection functions from both principals and labels to both confidentiality and integrity.
The $\pi_{P,C}$ and $\pi_{P,I}$ projections take principals and produce confidentiality and integrity, respectively,
and $\pi_{L,C}$ and $\pi_{L,I}$ do the same, but with labels as arguments.
We can think of $\pi_{P,C}(p)$ as ``the most confidential data that $p$ can read,'' while $\pi_{P,I}(p)$ is ``the highest integrity data that $p$ can write.''
We think of $\pi_{L,C}(\ell)$ as ``the confidentiality component of label $\ell$,'' while $\pi_{L,I}(\ell)$ is ``the integrity component of label $\ell$.''
With these functions, we can say that
$$\begin{array}{c}
  p \says*{\ell'} \canr(q, \ell) \leftrightarrow p \says*{\ell'} (\pi_{L,C}(\ell) \flowsto \pi_{P,C}(q)), \\[0.33em]
  \text{and} \\[0.33em]
  p \says*{\ell'}\canw(q,\ell) \leftrightarrow p \says*{\ell'} (\pi_{P,I}(q) \flowsto \pi_{L,I}(\ell)).
\end{array}$$
The reversal of the order here comes from the fact that integrity, as a flow ordering, is dual to confidentiality.


%% file: coq.tex
\section{Details of the Coq proofs}
\label{sec:coq-details}

In this appendix, we give some basic guidance to the Coq code, available at \\ \url{https://github.com/FLAFOL/flafol-coq}.

\paragraph{General Structure}
In the file \texttt{Term.v} we define the term language used by FLAFOL along with its type system. In the same file the module GroundInfo is defined. This module takes as parameters information necessary to instantiate the recipe specified in Section~\ref{sec:system-model}. For instance, it assumes the existence of a type of sorts and the existence of two sorts: Principal and Label. It also assumes that fresh variables can be generated and that equality of function and relation symbols is decidable. In the file \texttt{Formula.v} we define FLAFOL formulae.To simplify proofs, we use a locally nameless representation of variables \cite{locally-nameless-representation} and binding, and we prove some basic results about this binding discipline.
Note also that the definition of FLAFOL formulae is slightly different then that in the paper; rather than being part of the set $\mathcal{R}$, the permission relations are baked into the syntax of FLAFOL formulae directly.

We define the FLAFOL proof system in the file \texttt{Sequent.v}.
There are three ways in which our Coq formalism differs from the presentation of FLAFOL in Section~\ref{sec:proof-system}: (1) we use an equivalent presentation of the structural rules, (2) we use a slightly more general logic, and (3) we use two representations of the logic.

First, as is suggested by Pfenning \cite{structural-cut-elim}, we drop the structural rules from the logic (\textsc{Weakening}, \textsc{Exchange} and \textsc{Contraction}), modify our rules so that they never erase anything from the context and we prove that the removed rules are admissible.
This makes meta-theoretic proofs simpler.

Second, the logic described in the Coq is slightly more general than the one described in the paper.
In the Coq version the ground generalized principal has a label attached to it.
Originally we added ground-level labels to accommodate features that we left for future work, but we do not need them for this version of FLAFOL. 
To show that this is a generalization, for any FLAFOL proof without ground labels, we can simply assign the same ground label to every belief in the proof and acquire a valid proof in the Coq version.

Third, we have two representations of our logic.
The first is an (untyped) term language with the appropriate typing rules, and the second is a dependent inductive type.
The untyped version eases reasoning about equality, reduces compilation time, and makes proving the admissibility of weakening and substitution easier.
The typed version is easier to write automation tactics for. We have proved that both representations are equivalent.

\paragraph{Details of Cut Admissibility Proof}
In \texttt{NormalForm.v} we define a normal form for FLAFOL proofs.
The cut-elimination procedure uses normalization as an essential step.
\footnote{In the literature, ``normal proof'' refers to a cut-free proof, rather than a proof in FLAFOL's normal form.}
A proof is in normal form if all rules which do not manipulate formulae are higher in the proof tree than those which do. Formally, we define 2 normal forms, first and second normal form, which represent ``might use formula-manipulating rules'' and ``will not use formula manipulating rules'', respectively. A proof is in first normal form if, when a rule which manipulates something other than a formula is used, all subproofs above that rule are in second normal form, while a proof is in second normal form it if never uses any rules which manipulate formulae. 
The main result in this file is that every FLAFOL proof has a normal form.
\begin{thm}[FLAFOL Normal Form]
  If $\Gamma \proves \belief{\varphi}{g}$ is provable in FLAFOL, then it is provable with a proof in normal form.
\end{thm}

Lastly the file \texttt{Cut.v} contains the cut-elimination procedure.
First we normalize both proofs.
If they're both in First Normal Form but not in Second Normal Form, we proceed as Pfenning suggests in \cite{structural-cut-elim}: nested triple induction on the formula being cut and on both proofs.
If one of them is in Second Normal Form we use a different procedure.
This procedure consists of getting the dual rule to the last rule used in the proof that is in Second Normal Form (e.g. \textsc{VarL} for the \textsc{VarR} case) and make it the last rule to the other proof. Due to the covariant-contravariant nature of these rules and their duals, this is always possible.
For more details see lemmas Cut\_h1MCR and Cut\_h2MCR in \texttt{Cut.v}

\paragraph{Non-Interference}
In \texttt{Speaksfor.v} we define the relations $\spksfr$, $\influences$, define the function $\inmodals$ and prove Theorem~\ref{SFElim}.
The compatible supercontexts rules are defined in \texttt{CompatibleSuperContext.v}.
Finally, the Coq proof of Non-Interference is in \texttt{Noninterference.v}; it closely follows the pen-and-paper proof sketched in Section~\ref{sec:non-interference}.

\paragraph{Simulation}
The file \texttt{Simulation.v} contains the definition of the function $\odot$ a proof of the Simulation Theorem (Theorem~\ref{thm:simulation}).

%% file: fullproofsystem.tex
\section{The Full FLAFOL Proof System}
\label{sec:full-flafol-proof}

\begin{figure*}[p]
  \centering
  \pageOfRulesSize
  \begin{mathpar}
    \infer*[left=Ax]{ }{\Gamma, \belief{\varphi}{g} \proves \belief{\varphi}{g}} \and
    \infer*[left=Weakening]{\Gamma \proves \belief{\psi}{g}}{\Gamma, \belief{\varphi}{g'} \proves \belief{\psi}{g}}
    \\
    \infer*[left=Contraction]{\Gamma, (\belief{\varphi}{g}), (\belief{\varphi}{g}) \proves \belief{\psi}{g'}}{\Gamma, \belief{\varphi}{g} \proves \belief{\psi}{g'}} \and
    \infer*[left=Exchange]{\Gamma, (\belief{\varphi}{g_1}), (\belief{\psi}{g_2}), \Gamma' \proves \belief{\chi}{g}}{\Gamma, (\belief{\psi}{g_2}), (\belief{\varphi}{g_1}), \Gamma' \proves \belief{\chi}{g}}
    \\
    \infer*[left=FalseL]{ }{\Gamma, \belief{\False}{g} \proves \belief{\varphi}{g \cdot g'}} \and
    \infer*[left=TrueR]{ }{\Gamma \proves \belief{\True}{g}}
    \\
    \infer*[left=AndL]{\Gamma, (\belief{\varphi}{g}), (\belief{\psi}{g}) \proves \belief{\chi}{g'}}{\Gamma, (\belief{\varphi \land \psi}{g}) \proves \belief{\chi}{g'}} \and
    \infer*[left=AndR]{\Gamma \proves \belief{\varphi}{g} \\ \Gamma \proves \belief{\psi}{g}}{\Gamma \proves \belief{\varphi \land \psi}{g}}
    \\
    \infer*[left=OrL]{
      \Gamma, \belief{\varphi}{g} \proves \belief{\chi}{g'} \\
      \Gamma, \belief{\psi}{g} \proves \belief{\chi}{g'}}
    {\Gamma, (\belief{\varphi \lor \psi}{g}) \proves \belief{\chi}{g'}} \and
    \infer*[left=OrR1]{\Gamma \proves \belief{\varphi}{g}}{\Gamma \proves \belief{\varphi \lor \psi}{g}} \and
    \infer*[left=OrR2]{\Gamma \proves \belief{\psi}{g}}{\Gamma \proves \belief{\varphi \lor \psi}{g}}
    \\
    \infer*[left=ImpL]{
      \Gamma \proves \belief{\varphi}{\modalG{\ell}} \\
      \Gamma, \belief{\psi}{g} \proves \belief{\chi}{g'} \\
    }{\Gamma, (\belief{\varphi \mimpl* \psi}{g}) \proves \belief{\chi}{g'}} \and
    \infer*[left=ImpR]{\Gamma, \belief{\varphi}{\modalG{\ell}} \proves \belief{\psi}{g}}{\Gamma \proves \belief{\varphi \mimpl* \psi}{g}}
    \\
    \infer*[left=ForallL]{\Gamma, \belief{\subst{\varphi}{x}{t}}{g} \proves \belief{\psi}{g'}}{\Gamma, (\belief{\forallexp*{\varphi}}{g}) \proves \belief{\psi}{g'}} \and
    \infer*[left=ForallR]{\Gamma \proves \belief{\varphi}{g} \\ x \notin \fv(\Gamma,g)}{\Gamma \proves \belief{\forallexp*{\varphi}}{g}}
    \\
    \infer*[left=ExistsL]{\Gamma, \belief{\varphi}{g} \proves \belief{\psi}{g'} \\ x \notin \fv(\Gamma, \psi, g, g')}{\Gamma, (\belief{\existsexp*{\varphi}}{g}) \proves \belief{\psi}{g'}} \and
    \infer*[left=ExistsR]{\Gamma \proves \belief{\subst{\varphi}{x}{t}}{g}}{\Gamma \proves \belief{\existsexp*{\psi}}{g}}
    \\
    \infer*[left=SaysL]{\Gamma, \belief{\varphi}{g \cdot \modal{p}{\ell}} \proves \belief{\psi}{g'}}{\Gamma, \belief{p \says*{\ell} \varphi}{g} \proves \belief{\psi}{g'}} \and
    \infer*[left=SaysR]{\Gamma \proves \belief{\varphi}{g \cdot \modal{p}{\ell}}}{\Gamma \proves \belief{p \says*{\ell} \varphi}{g}}
    \\
    {\mprset{fraction={===}}
    \infer*[left=SelfL]{\Gamma, (\belief{\varphi}{g \cdot \modal{p}{\ell} \cdot g'}) \proves \belief{\psi}{g''}}{\Gamma, (\belief{\varphi}{g \cdot \modal{p}{\ell} \cdot \modal{p}{\ell} \cdot g'}) \proves \belief{\psi}{g''}} \and
    \infer*[left=SelfR]{\Gamma \proves \belief{\varphi}{g \cdot \modal{p}{\ell} \cdot g'}}{\Gamma \proves \belief{\varphi}{g \cdot \modal{p}{\ell} \cdot \modal{p}{\ell} \cdot g'}}
    }
    \\
    \infer*[left=VarL]{
      \Gamma, (\belief{\varphi}{g \cdot \modal{p}{\ell'} \cdot g'}) \proves \belief{\psi}{g''} \\\\
      \Gamma, (\belief{\varphi}{g \cdot \modal{p}{\ell} \cdot g'}) \proves \belief{\ell \flowsto \ell'}{g \cdot \modal{p}{\ell'}}}
    {\Gamma, (\belief{\varphi}{g \cdot \modal{p}{\ell} \cdot g'}) \proves \belief{\psi}{g''}} \and
    \infer*[left=VarR]{
      \Gamma \proves \belief{\varphi}{g \cdot \modal{p}{\ell'} \cdot g'} \\
      \Gamma \proves \belief{\ell' \flowsto \ell}{g \cdot \modal{p}{\ell}}}
    {\Gamma \proves \belief{\varphi}{g \cdot \modal{p}{\ell} \cdot g'}}
    \\
    \infer*[left=FwdL]{
      \Gamma, (\belief{\varphi}{g \cdot \modal{q}{\ell} \cdot g'}) \proves \belief{\chi}{g''}\\\\
      \Gamma, (\belief{\varphi}{g \cdot \modal{p}{\ell} \cdot g'}) \proves \belief{\canr(q, \ell)}{g \cdot \modal{p}{\ell}}\\\\
      \Gamma, (\belief{\varphi}{g \cdot \modal{p}{\ell} \cdot g'}) \proves \belief{\canw(p, \ell)}{g \cdot \modal{q}{\ell}}}
    {\Gamma, \belief{\varphi}{g \cdot \modal{p}{\ell} \cdot g' \proves \belief{\chi}{g''}}} \and
    \infer*[left=FwdR]{
      \Gamma \proves \belief{\varphi}{g \cdot \modal{p}{\ell} \cdot g'}\\\\
      \Gamma \proves \belief{\canr(q, \ell)}{g \cdot \modal{p}{\ell}}\\\\
      \Gamma \proves \belief{\canw(p, \ell)}{g \cdot \modal{q}{\ell}}}
    {\Gamma \proves \belief{\varphi}{g \cdot \modal{q}{\ell} \cdot g'}}
    \\
    \infer*[left=FlowsToRefl]{ }{\Gamma \proves \belief{\ell \sqsubseteq \ell}{g}} \and
    \infer*[left=FlowsToTrans]{\Gamma \proves \belief{\ell_1 \sqsubseteq \ell_2}{g} \\ \Gamma \proves \belief{\ell_2 \sqsubseteq \ell_3}{g}}{\Gamma \proves \belief{\ell_1 \sqsubseteq \ell_3}{g}} \and
    \infer*[left=CRVar]{
      \Gamma \proves \belief{\canr(p, \ell_2)}{g}\\
      \Gamma \proves \belief{\ell_1 \sqsubseteq \ell_2}{g}}
    {\Gamma \proves \belief{\canr(p, \ell_1)}{g}} \and
    \infer*[left=CWVar]{
      \Gamma \proves \belief{\canw(p, \ell_2)}{g}\\
      \Gamma \proves \belief{\ell_2 \sqsubseteq \ell_1}{g}}
    {\Gamma \proves \belief{\canw(p, \ell_1)}{g}}
  \end{mathpar}
  \caption{Full FLAFOL Proof System}
  \label{fig:full-proof-rules}
\end{figure*}

The full FLAFOL proof system can be found in Figure~\ref{fig:full-proof-rules}.


%% file: CompatibleSupercontexts.tex
\section{Compatible Supercontexts}
\label{sec:comp-superc}

\begin{figure*}[p]
  \centering
  \pageOfRulesSize
  \begin{mathpar}
    \infer*[left=CSCRefl]{ }{\Gamma \ll  \Gamma \proves  \belief{\varphi}{g}} \and
    \infer*[left=CSCUnion]{\Delta_1 \ll  \Gamma \proves \belief{\varphi}{g}\\ \Delta_2 \ll \Gamma \vdash \belief{\varphi}{g}}{\Delta_1 \cup \Delta_2 \ll \Gamma \vdash \belief{\varphi}{g}} \and
    \infer*[left=CSCContraction]{\Delta \ll \Gamma, (\belief{\varphi}{g}), (\belief{\varphi}{g}) \proves \belief{\psi}{g'}}{\Delta \ll \Gamma, \belief{\varphi}{g} \proves \belief{\psi}{g'}} \and
    \infer*[left=CSCExchange]{\Delta \ll \Gamma, (\belief{\varphi}{g_1}), (\belief{\psi}{g_2}), \Gamma' \proves \belief{\chi}{g}}{\Delta \ll \Gamma, (\belief{\psi}{g_2}), (\belief{\varphi}{g_1}), \Gamma' \proves \belief{\chi}{g}} \and
    \infer*[left=CSCAndL]{\Delta \ll \Gamma, (\belief{\varphi}{g}), (\belief{\psi}{g}) \proves \belief{\chi}{g'}}{\Delta \ll \Gamma, (\belief{\varphi \land \psi}{g}) \proves \belief{\chi}{g'}} \and
    \infer*[left=CSCAndR1]{\Delta \ll \Gamma \vdash \belief{\varphi}{g}}{\Delta \ll \Gamma \vdash \belief{\varphi \land \psi}{g}} \and
    \infer*[left=CSCAndR2]{\Delta \ll \Gamma \vdash \belief{\psi}{g}}{\Delta \ll \Gamma \vdash \belief{\varphi \land \psi}{g}} \and
    \infer*[left=CSCOrL1]{\Delta \ll \Gamma, \belief{\varphi}{g} \proves \belief{\chi}{g'}}{\Delta \ll \Gamma, (\belief{\varphi \lor \psi}{g}) \proves \belief{\chi}{g'}} \and
    \infer*[left=CSCOrL2]{\Delta \ll \Gamma, \belief{\psi}{g} \proves \belief{\chi}{g'}}{\Delta \ll \Gamma, (\belief{\varphi \lor \psi}{g}) \proves \belief{\chi}{g'}} \and
    \infer*[left=CSCOrR1]{\Delta \ll \Gamma \proves \belief{\varphi}{g}}{\Delta \ll \Gamma \proves \belief{\varphi \lor \psi}{g}} \and
    \infer*[left=CSCOrR2]{\Delta \ll \Gamma \proves \belief{\psi}{g}}{\Delta \ll \Gamma \proves \belief{\varphi \lor \psi}{g}} \and
    \infer*[left=CSCImpL1]{\Delta \ll \Gamma, \belief{\psi}{g} \proves \belief{\chi}{g'}}{\Delta \ll \Gamma, (\belief{\varphi \mimpl* \psi}{g}) \proves \belief{\chi}{g'}} \and
    \infer*[left=CSCImpL2]{\Delta \ll \Gamma \proves \belief{\varphi}{\modalG{}}}{\Delta \ll \Gamma, (\belief{\varphi \mimpl* \psi}{g}) \proves \belief{\chi}{g'}} \and
    \infer*[left=CSCImpR]{\Delta \ll \Gamma, \belief{\varphi}{\modalG{}} \proves \belief{\psi}{g}}{\Delta \ll \Gamma \proves \belief{\varphi \mimpl* \psi}{g}} \and
    \infer*[left=CSCForallL]{\Delta \ll \Gamma, \belief{\subst{\varphi}{x}{t}}{g} \proves \belief{\psi}{g'}}{\Delta \ll \Gamma, (\belief{\forallexp*{\varphi}}{g}) \proves \belief{\psi}{g'}} \and
    \infer*[left=CSCForallR]{\Delta \ll \Gamma \proves \belief{\varphi}{g} \\ x \notin \fv(\Gamma, g)}{\Delta \ll \Gamma \proves \belief{\forallexp*{\varphi}}{g}} \and
    \infer*[left=CSCExistsL]{\Delta \ll \Gamma, \belief{\varphi}{g} \proves \belief{\psi}{g'} \\ x \notin \fv(\Gamma, \psi, g, g')}{\Delta \ll \Gamma, (\belief{\existsexp*{\varphi}}{g}) \proves \belief{\psi}{g'}} \and
    \infer*[left=CSCExistsR]{\Delta \ll \Gamma \proves \belief{\subst{\varphi}{x}{t}}{g}}{\Delta \ll \Gamma \proves \belief{\existsexp*{\varphi}}{g}} \and
    \infer*[left=CSCSaysL]{\Delta \ll \Gamma, \belief{\varphi}{g \cdot \modal{p}{\ell}} \proves \belief{\psi}{g'}}{\Delta \ll \Gamma, \belief{p \says*{\ell} \varphi}{g} \proves \belief{\psi}{g'}} \and
    \infer*[left=CSCSaysR]{\Delta \ll \Gamma \proves \belief{\varphi}{g \cdot \modal{p}{\ell}}}{\Delta \ll \Gamma \proves \belief{p \says*{\ell} \varphi}{g}} \and
    {\mprset{fraction={===}}
    \infer*[left=CSCSelfL]{\Delta \ll \Gamma, (\belief{\varphi}{g \cdot \modal{p}{\ell} \cdot g'}) \proves \belief{\psi}{g''}}{\Delta \ll \Gamma, (\belief{\varphi}{g \cdot \modal{p}{\ell} \cdot \modal{p}{\ell} \cdot g'}) \proves \belief{\psi}{g''}} \and
    \infer*[left=CSCSelfR]{\Delta \ll \Gamma \proves \belief{\varphi}{g \cdot \modal{p}{\ell} \cdot g'}}{\Delta \ll \Gamma \proves \belief{\varphi}{g \cdot \modal{p}{\ell} \cdot \modal{p}{\ell} \cdot g'}}
    } \and
    \infer*[left=CSCVarL]{
      \Delta \ll \Gamma, (\belief{\varphi}{g \cdot \modal{p}{\ell'} \cdot g'}) \proves \belief{\psi}{g''} \\
      \Gamma, (\belief{\varphi}{g \cdot \modal{p}{\ell} \cdot g'}) \proves \belief{\ell \flowsto \ell'}{g \cdot \modal{p}{\ell'}}}
    {\Delta \ll \Gamma, (\belief{\varphi}{g \cdot \modal{p}{\ell} \cdot g'}) \proves \belief{\psi}{g''}} \and
    \infer*[left=CSCVarR]{
      \Delta \ll \Gamma \proves \belief{\varphi}{g \cdot \modal{p}{\ell'} \cdot g'} \\
      \Gamma \proves \belief{\ell' \flowsto \ell}{g \cdot \modal{p}{\ell}}}
    {\Delta \ll \Gamma \proves \belief{\varphi}{g \cdot \modal{p}{\ell} \cdot g'}} \and
    \infer*[left=CSCFwdL]{
      \Delta \ll \Gamma, (\belief{\varphi}{g \cdot \modal{q}{\ell} \cdot g'}) \proves \belief{\chi}{g''}\\
      \Gamma, (\belief{\varphi}{g \cdot \modal{p}{\ell} \cdot g'}) \proves \belief{\canr(q, \ell)}{g \cdot \modal{p}{\ell}}\\
      \Gamma, (\belief{\varphi}{g \cdot \modal{p}{\ell} \cdot g'}) \proves \belief{\canw(p, \ell)}{g \cdot \modal{q}{\ell}}}
    {\Delta \ll \Gamma, \belief{\varphi}{g \cdot \modal{p}{\ell} \cdot g' \proves \belief{\chi}{g''}}} \and
    \infer*[left=CSCFwdR]{
      \Delta \ll \Gamma \proves \belief{\varphi}{g \cdot \modal{p}{\ell} \cdot g'}\\
      \Gamma \proves \belief{\canr(q, \ell)}{g \cdot \modal{p}{\ell}}\\
      \Gamma \proves \belief{\canw(p, \ell)}{g \cdot \modal{q}{\ell}}}
    {\Delta \ll \Gamma \proves \belief{\varphi}{g \cdot \modal{q}{\ell} \cdot g'}}
  \end{mathpar}
  \caption{Compatible Supercontext Rules}
  \label{fig:compat-superctxt-rules}
\end{figure*}

Figure~\ref{fig:compat-superctxt-rules} contains the full rules for compatible super-contexts.
